\documentclass[usenatbib]{mn2e}
\input{epsf}

\def\lsim{ \lower .75ex \hbox{$\sim$} \llap{\raise .27ex \hbox{$<$}} }
\def\gsim{ \lower .75ex \hbox{$\sim$} \llap{\raise .27ex \hbox{$>$}} }
\newcommand{\apjl}{ApJL}
\newcommand{\apj}{ApJ}
\newcommand{\araa}{ARA\&A}
\newcommand{\aj}{AJ}
\newcommand{\mnras}{MNRAS}

\newcommand{\apjs}{ApJS}

\newcommand{\aap}{A\&A}

\newcommand{\pasj}{PASJ}
\newcommand{\nat}{Nature}

\begin{document}
 
\title[EROs in a hierarchical universe]
{Massive, red galaxies in a hierarchical universe-I 
Counts of  Extremely Red Objects and basic properties}

\author[Gonzalez-Perez  et al.]{
\parbox[t]{\textwidth}{
\vspace{-1.0cm}
V.\,Gonzalez-Perez$^{1}$,
C.\,M.\,Baugh$^{2}$,
C.G.Lacey$^{2}$,
C. Almeida$^{2}$.
}
\\
$^{1}$Institut de Ci\`{e}ncies de l'Espai (CSIC/IEEC), F. de Ciencies, Torre C5 Par 2a, UAB, Bellaterra, 08193 Barcelona, Spain\\
$^{2}$Institute for Computational Cosmology, Department of Physics, 
University of Durham, South Road, Durham, DH1 3LE, UK.
}
 
\maketitle
\begin{abstract}
We present predictions for the abundance and nature of Extremely Red Objects 
(EROs) in the $\Lambda$ cold dark matter model. EROs are red, massive galaxies 
observed at $z\geq 1$ and their numbers and properties pose a challenge 
to hierarchical galaxy formation models. We compare the predictions from 
two published models, one of which invokes a ``superwind'' to regulate 
star formation in massive haloes 
and the other which suppresses gas cooling in haloes through ``radio-mode'' AGN feedback. 
The superwind model underestimates the number counts of EROs by an 
order of magnitude, whereas the radio-mode AGN feedback model gives excellent 
agreement with the number counts and redshift distribution of EROs. 
In the AGN feedback model the ERO population is dominated by 
old, passively evolving galaxies, whereas observations favour an equal 
split between old galaxies and dusty starbursts. Also, the model predicts 
a more extended redshift distribution of passive galaxies than is observed. 
These comparisons suggest that star formation may be quenched too efficiently 
in this model. 
\end{abstract}

\section{Introduction }
\defcitealias{baugh05}{\citeauthor{baugh05}} 	
\defcitealias{bower06}{\citeauthor{bower06}}

Advances in near-infrared detector technology have made possible the 
identification of large samples of massive, red galaxies at redshifts 
when the universe was less than half its current age 
\citep[e.g.][]{elston88,thompson99,daddi00,mccarthy01,cimatti02a,smith02,
simpson06,conselice08}. 
These objects, according to estimates based on their K-band fluxes, have stellar masses 
comparable to the most massive galaxies in place today 
\citep[][]{cimatti04,glazebrook04}. Extremely red objects (EROs) are among these galaxies. Their extremely 
red optical-near infrared colours, indicate either a star-forming galaxy 
with heavy obscuration or an aged stellar population with little recent 
star formation \citep[][]{pozzetti00,smail02,cimatti03}. 
In the latter case, the implied age of the stellar population is 
uncomfortably close to the age of the Universe at the redshift of 
observation \citep[][]{cimatti02a,smith02}.

At face value either ERO scenario might appear difficult 
to reconcile with a universe in which structure in the dark matter 
grows in a bottom-up sequence, with smaller haloes merging to make 
more massive haloes. If the universe is assumed to be populated 
by low mass dark matter haloes at high redshift, and, as is traditionally argued, 
these haloes are ineffective at making galaxies due to feedback from supernova explosions, 
how can the stars in EROs have formed by the high redshifts suggested 
by observations? Where are the deep potential wells in which enough 
cold gas has accumulated for luminous, dusty starbursts to take place? 
In practice, two effects act to make it possible for the stars in massive 
objects to form at high redshift in a cold dark matter model. First, 
the range of halo masses collapsing at a given redshift is actually 
very broad in the CDM model, due to the shape of the power spectrum of density 
fluctuations \citep[see e.g.][]{mo02}. 
Second, the progenitors of massive haloes form at an earlier epoch than 
similar mass progenitors which end up in less extreme mass haloes. This in turn 
means that these progenitors can start to form stars earlier \citep[see][]{neistein06, delucia06}. 

Earlier generations of hierarchical galaxy formation models did indeed 
fail to match the observed abundance of EROs, 
by around an order of magnitude \citep[][]{smith02,firth02,roche02} and 
predicted an ERO redshift distribution that was too shallow \citep[][]{cimatti02b,somerville04}. 
This initial 
failure led to attempts to resurrect the monolithic collapse scenario for 
massive galaxy formation, in which all the stars in a galaxy form at the 
same epoch \citep{eggen62}. 
Pure luminosity evolution (PLE) models, simple parametric functions which are meant 
to describe the brightness of a stellar population as a function of time, can 
be adjusted to give reasonable matches to the number counts and redshift 
distributions of EROs \citep[e.g.][]{daddi00}. However, such an approach 
soon falls apart when applied to more than one type of galaxy. \citet{manfred06} demonstrated that PLE models that give mass-to-light ratios 
consistent with the red sequence of galaxies found in clusters 
have the drawback that they overpredict the number counts of 
magnitude-limited samples at faint magnitudes.  

Recently there has been much development in the modelling of the formation 
of massive galaxies. This activity was primarily directed at solving the 
problem of matching the bright end of the local galaxy luminosity function, 
but also has implications for the abundance of massive galaxies at high 
redshift. Typically, hierarchical models, based on the current best 
fitting values of the cosmological parameters, produce too many massive 
galaxies at the present day unless some physical process is invoked 
to restrict gas cooling in massive haloes \citep[see e.g.][for a review of 
plausible mechanisms]{benson03}. The most promising candidates are the 
heating of gas in quasi-hydrostatically cooling haloes in 
which ``radio mode'' AGN feedback is considered to be effective 
\citep{bower06,croton06,kang06,cattaneo06,lagos08}, the ejection of gas from 
intermediate mass haloes in a wind, which could be driven by supernovae 
\citep[e.g.][]{benson03,baugh05} 
or ``quasar mode'' feedback which follows accretion of cold gas 
driven by a galaxy merger or the collapse of an unstable disk \citep{granato04,hopkins06,menci06}. Recent models 
\citep[see][]{debora07,pg07,somer08} incoporate both the radio and quasar modes 
of feedback and their corresponding impact on star formation in the host galaxy.

Several authors have gone back to the problem of the abundance of massive 
galaxies following the recent improvements in the theoretical models. 
There are several conditions to bear in mind when assessing the merits of 
a model which claims to reproduce a given dataset, such as the number 
counts of EROs: 
(i) Does the model attempt to follow the full galaxy population or 
just some component of it, such as spheroids? 
(ii) Does the model follow the galaxy population to the present day? 
Can the same model, which successfully matches observations at high 
redshift, also reproduce local data? 
(iii) How is the comparison with observations carried out? Is it direct, 
relying on photometric selection of galaxies or is it indirect, using 
quantities which are not directly observed, such as stellar mass, 
to compare galaxies? 

\citet{nagamine05} compared calculations made with Eulerian and 
Lagrangian gas-dynamics codes and found that, with their choice of 
``sub-grid" physics to describe baryonic processes, they did not 
have a problem in reproducing the number of massive galaxies 
seen at high redshift. However, producing the observed number of 
galaxies with red colours proved to be more challenging. \citeauthor{nagamine05} 
found they could obtain the correct number of EROs if they assumed all 
EROs to be heavily extincted starbursts (applying by hand an extinction equivalent to 
$E(B-V)=0.4$ to each galaxy), and without any galaxies which displayed the colours of 
old, passively evolving stellar populations. Observationally, 
passive galaxies and dusty starbursts contribute roughly equally to the counts 
of EROs \citep[e.g.][]{mannucci02,cimatti02a}. 

Models in which radio mode AGN feedback suppresses gas cooling in massive haloes 
seem to have taken significant steps towards solving the massive red 
galaxy problem. \citet{manfred07} studied the abundance of 
massive galaxies in the semi-analytical model of \citet{delucia07}. 
This model actually overpredicts the number counts of galaxies faintwards of 
$K_{AB} \sim 20$ and predicts more extended redshift distributions than 
are observed, which means that the model has {\it too much} star formation at 
high redshift.  \citet{kang06}, on the other hand, claim to 
reproduce the number of EROs with $(R-K)>5$, but not the number defined 
by redder cuts. The \citeauthor{kang06} model, however, overpredicts the number 
of bright galaxies in the K-band luminosity function today. Another 
problem is that this model predicts a lower median redshift for K-selected galaxies 
than is observed.

Alternative models have been proposed in which feedback in massive haloes 
occurs during the quasar mode of mass accretion onto black holes, 
i.e. the QSO phase of an AGN. The QSO episode is assumed to be 
triggered by a galaxy interaction or merger, or the dynamical instability of 
a galactic disk. As a result cold gas moves towards the centre of the galaxy, 
and some fraction is added to a supermassive black hole via an accretion disk. 
The QSO activity peaks at high redshift, when the merger rate is higher and when 
larger reservoirs of cold gas are in place. At the onset of the QSO phase, the 
remaining gas in the galaxy is blasted out and the star formation 
terminates \citep{silk98}.  
This quasar mode feedback is distinct from the radio mode mentioned above 
which acts predominantly at later times or in haloes which are massive enough for 
quasi-static hot atmospheres to form. In the radio mode, the black hole is fuelled 
by the accretion of gas in a cooling flow. Quasar mode feedback has been 
implemented into several semi-analytical models. \citet{granato04} argued that 
QSO feedback eventually truncates star formation, causing the stellar 
population of the galaxy to age, resulting in EROs. However, these authors were 
only able to track galactic spheroids in their model.
\citet{fontanot07} consider a semi-analytical code with a new 
cooling model which is able to match the counts but not the redshifts 
of sub-mm  galaxies using a standard stellar initial mass function. 
However, their preferred model overpredicts the local abundance of 
bright galaxies by at least an order of magnitude.
Perhaps the most successful model of this type to date is that of 
\citet{menci06} which is able to reproduce 
the counts and redshift distribution of EROs, along with the rest-frame 
B-band luminosity function over the interval $0.4\leq z \leq 3.5$. 
With QSO feedback models, the problem of overcooling in massive haloes 
at low redshifts remains, unless the gas expelled in the QSO feedback episode is 
prevented from being recaptured. As mentioned above, recent models have 
incorporated both radio and quasar modes of feedback \citep{debora07,somer08}.

In this paper we examine the predictions for EROs from two published 
semi-analytical models,  \citet{baugh05} and \citet{bower06}.
The models invoke different mechanisms to suppress star formation in 
massive galaxies. In \citeauthor{baugh05}, a wind ejects gas from low and intermediate 
mass haloes, with the mass ejection rate tied to the star formation rate. 
Due to the ejection of gas, the baryon fraction in massive haloes is 
therefore lower than the universal baryon fraction, reducing the cooling 
rate. The \citeauthor{bower06} model invokes radio-mode AGN feedback to suppress 
gas cooling in massive haloes. The parameters in both models are set by the 
requirement that they reproduce a subset of local galaxy observations. 
\citet{almeida07lrg} investigated the predictions of these models for 
the abundance of luminous red galaxies \citep{eisenstein01}. Here, 
we extend this comparison to much higher redshifts by looking at the 
predictions for EROs. This is the first in a series of 
three papers. In this paper we look at the number counts, redshift distributions and 
other basic properties of EROs in the models; in the second paper we 
study the clustering of EROS and their descendants and in the third paper we 
compare the implications of different selections for red galaxies. 
This paper is organized as follows. In Section~\ref{sec:model}, we summarize 
the two galaxy formation models used to study the EROs population. 
The predictions for the abundance and redshift distribution of EROs 
are given in Section~3. The nature of EROs in the model is discussed in 
Section~4 and their basic properties are presented in Section~5. 
Our conclusions are presented in Section~\ref{sec:conclusions}.

The bands used here correspond to the R band from {\sc SUBARU}, 
centred at $0.65$ $\mu m$, and  the K band from {\sc UKIRT}, 
with a central wavelength of $2.2$ $\mu m$. All magnitudes used 
in this paper are on the Vega system, unless otherwise specified. 
The cosmological parameters of the models are given in Section~\ref{sec:model}. 

\section{Galaxy formation model}\label{sec:model}

We predict the abundance and properties of EROs in a $\Lambda$CDM 
universe using the {\tt GALFORM} semi-analytical galaxy formation 
code developed by \citet{cole00}, and extended by \citet{benson03} and \citet{bower06}. Semi-analytical models 
use simple, physically motivated recipes and rules to follow the 
fate of baryons in a universe in which structures grow hierarchically 
through gravitational instability \citep[see][for an introduction to  
hierarchical galaxy formation models]{baugh06}. 

In this paper we focus our attention on two published models, 
\citet{baugh05} and \citet{bower06}. The parameters of these models were 
fixed with reference to a subset of the available observations of galaxies, 
mostly at low redshift. In this paper we extract predictions for the number 
and nature of ERO galaxies without adjusting the values of any of the model 
parameters. Although none of the datasets used to set the model parameters 
explicitly referred to EROs, one of our priorities in adjusting parameter 
values is to obtain as good a match as possible to the bright 
end of the local field galaxy luminosity function. Observationally, the 
bright end of the luminosity function tends to be dominated by galaxies 
with red colours and passively evolving stellar populations \citep[e.g.][]{norberg02}. Hence by reproducing the observed luminosity function, the models 
have approximately the right number of bright red galaxies today. By testing 
the predictions for EROs, we are therefore probing the evolution 
of the bright red galaxy population in the models to $z > 1$.  

In addition to reproducing local galaxy data, the models have some notable 
successes at high redshift. The \citeauthor{bower06} model matches the 
inferred evolution of the stellar mass function to $z=4.5$. The 
\citeauthor{baugh05} model matches the number and redshift distribution 
of galaxies detected by their emission at sub-millimetre wavelengths 
\citep[see also the predictions presented by][]{lacey08}, the luminosity 
function of Lyman break galaxies and the abundance and clustering of 
Lyman-alpha emitters \citep{morganI,morganII,alvaro}. 

We now recap some of the key features of the models for the study presented 
in this paper and draw attention to places where the two models differ. A 
similar comparison of the two models can be found in Almeida et~al. (2007, 2008). 
For a comprehensive inventory of the ingredients of the models, we refer the 
reader to the original papers \citep[see also][for a recap of the
ingredients of \citeauthor{baugh05}]{lacey08}. 

\begin{itemize} 
\item{\it Cosmology.} \citeauthor{baugh05} use the canonical ($\Lambda$CDM) 
parameters: matter density, $\Omega_{0}=0.3$, cosmological constant, 
$\Lambda_{0} = 0.7$, baryon density, $\Omega_{b}=0.04$, a normalization of 
density fluctuations given by $\sigma_{8}=0.93$ and a Hubble constant $h=0.7$ 
in units of 100 km s$^{-1}$ Mpc$^{-1}$. \citeauthor{bower06} adopt the 
cosmological parameters of the Millennium Simulation \citep{springel05}, 
which are in better agreement with recent constraints from measurements 
of the cosmic microwave background radiation and large scale galaxy clustering 
\citep[e.g.][]{sanchez06}: $\Omega_{0}=0.25$, $\Lambda_{0} = 0.75$, 
$\Omega_{b}=0.045$, $\sigma_{8}=0.9$ and $h=0.73$. 

\item {\it Dark matter halo merger trees.} 
The \citeauthor{baugh05} model employs merger trees generated using 
the Monte Carlo algorithm introduced by \citet{cole00}. The \citeauthor{bower06} 
model uses halo merger histories extracted from the Millennium Simulation 
\citep{springel05}. These two approaches have been shown to yield similar 
results for galaxies brighter than a limiting faint magnitude which is determined 
by the mass resolution of the N-body trees \citep{helly03}. In the case of 
the Millennium merger trees, this limit is several magnitudes fainter than 
$L_{*}$ and so will have little consequence for the study of EROs, which are 
much brighter. We compute number counts in the \citeauthor{baugh05} model by 
growing merger histories for representative grids of halo masses laid down 
at a range of output redshifts, rather than outputting the branches of one 
set of trees grown from $z=0$.  In this way, we avoid any 
biases in the progenitor distribution which may develop over large lookback 
times. Recently, a more accurate Monte Carlo prescription for generating merger 
histories has been developed and it would be instructive to re-run 
the \citeauthor{baugh05} model with these modified trees \citep[][]{parkinson08}.

\item{\it Feedback processes.}
Both models regulate star formation by the injection of energy from supernova 
explosions into the cold gas reservoir. In \citeauthor{baugh05}, there are two consequences of 
this energy injection which are parameterized in different ways and which also 
differ in the fate of the reheated gas. In the ``standard'' mode of supernova 
feedback, cold gas is heated and ejected from the galactic disk. In \citeauthor{baugh05}, this 
gas is not allowed to recool until a new halo has formed (as signalled by a 
doubling of the halo mass), whereupon this reheated gas is incorporated into 
the hot gas atmosphere of the new halo. In the ``superwind'' mode of feedback, 
reheated gas is expelled completely from the halo, and in this particular 
implementation is never allowed to recool. These two modes of supernova feedback 
can operate side-by-side in a given halo, with relative strengths determined  
by the parameter values chosen \citep[see][]{lacey08}. \citet{benson03} 
give a more detailed discussion of these modes of feedback. The superwind suppresses 
the formation of bright galaxies in massive haloes. It achieves this by 
ejecting gas from the progenitors of the massive halo, so that the hot 
gas reservoir is depleted, and the massive halo effectively has a baryon fraction 
that is lower than the universal value. One concern is that the parameters in the 
feedback recipes are chosen without reference to the total amount of energy 
available from supernova explosions. Once a reasonable match to the bright end 
of the luminosity function has been obtained, the implied efficiency with which 
the energy released by supernovae must couple to heating the intergalactic medium is uncomfortably 
high. \citeauthor{bower06} also invoke ``standard'' supernovae feedback, though with very different 
parameters values than those used in \citeauthor{baugh05}. This is due in part to a modification in the 
handling of the gas reheated by supernova feedback. Rather than being placed in limbo until a new halo 
forms, the reheated gas is incorporated into the hot halo after some number of 
halo dynamical times. Hence gas can recool more rapidly in \citeauthor{bower06}, which explains why 
the standard supernova feedback parameters are set to values which correspond to 
stronger feedback than in \citeauthor{baugh05} In \citeauthor{bower06}, there is no superwind feedback. The formation of bright 
galaxies is suppressed by staunching the cooling flow in massive haloes using ``radio-mode'' 
AGN feedback. This is 
achieved by the injection of energy into the quasistatic hot halo, which is generated  
by the accretion of matter onto a central supermassive black hole \citep[see][for a description of the model of black hole growth]{malbon07}.

\item{\it Star formation.}: 
In the models, stars can form quiescently in galactic disks or in bursts. The two 
models adopt different redshift dependencies for the time-scale for quiescent star 
formation. \citeauthor{baugh05} adopt a fixed time-scale whereas in \citeauthor{bower06}, the time-scale depends on the 
dynamical time. In both models, mergers can trigger starbursts, though the conditions 
for a burst to occur are different. \citeauthor{baugh05} include only bursts triggered by galaxy mergers, 
whereas \citeauthor{bower06} also consider bursts which result from disks 
becoming dynamically unstable to the formation of a bar.

\item{\it Stellar Initial Mass Function (IMF).}
Both models adopt the \citet{kennicutt_imf} IMF for quiescent star formation.
\citeauthor{bower06} also use this IMF in starbursts, whereas \citeauthor{baugh05} 
invoke a top-heavy IMF. This choice, though controversial, is the key to the 
success of \citeauthor{baugh05} in reproducing sub-mm galaxies number counts and also the metallicity 
of the intracluster medium \citep{nagashima05}. The yield of metals 
and the fraction of gas recycled in star formation are determined by the choice of 
IMF. 

\end{itemize}

Finally in this section, we discuss one element which the models have in common, 
the treatment of dust extinction, which is important for our purposes as it has 
an impact on galaxy colours. Both models employ an extension of the dust extinction 
calculation introduced by \citet{cole00}. {\tt GALFORM} makes a self-consistent calculation 
of the dust optical depth of a galaxy, computing the gas mass and metallicity based on 
a chemical evolution model and making a prediction for the scale length of the 
disk and bulge components. The size calculation is also explained 
by \citeauthor{cole00} (see Almeida et~al. 2007 for a 
test of this calculation for spheroids). In brief, for the disk component, the model 
assumes the conservation of angular momentum of the gas. The size of spheroids is calculated 
following a merger by considering the conservation of energy and the application of 
the virial theorem. The size of the disk and bulge components also takes into account 
the gravity of the baryons and the dark matter. The stars are assumed to be mixed in 
with the dust, with the possibility that the two components may have different scale 
heights. The dust is assumed to have the properties consistent with the extinction law 
observed in the Milky Way. An inclination angle is assigned at random to the galactic disk. 
The dust extinction is then computed using the results of radiative transfer calculations 
carried out by Ferrara et~al. (1999). This model is a significant improvement over 
calculations using foreground screens, in which a slab of dust is assumed to be along 
the line of sight to the stars and an empirical estimate is made of the optical depth. 

The extension we apply to the dust extinction model of \citeauthor{cole00} is to assume 
that some fraction of the dust is in the form of dense molecular clouds where the stars 
form (see Baugh et~al. 2005). This modifies the extinction of starlight, particularly at 
very short wavelengths. Emission in the ultra-violet is dominated by hot, massive stars 
which have short lifetimes. The massive stars spend a significant fraction of their lifetime 
inside a molecular cloud, depending upon the time-scale adopted for the star to escape from 
the cloud, which is a parameter of the dust model \citep{granato00}. This hybrid scheme with 
diffuse and molecular cloud dust components mimics the more rigorous calculation of dust 
extinction carried out by the spectro-photometric code {\tt GRASIL} \citep{silva98}.

\section{The counts of EROs}\label{sec:results}

Here we compare the abundance and redshift distribution of EROs and K-selected 
samples predicted by the two models with observations. 
We begin by considering how well the models reproduce the number 
counts of K-selected galaxies (\S3.1), before presenting predictions for the surface 
density of EROs (\S3.2). Redshift distributions are discussed in \S\ref{sec:nz}. 

\subsection{Total K-band number counts} 
\label{kcounts}

\begin{figure}
{\epsfxsize=8.5truecm
\epsfbox[26 3 570 540]{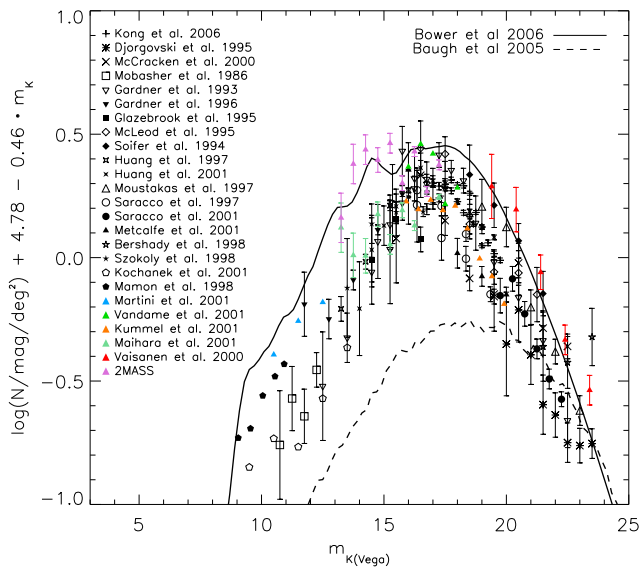}}
\caption{
The differential K-band number counts, plotted after dividing by a power law 
(as given in the label on the y-axis) which is the one that best fits the observed counts. 
The solid line shows the counts predicted by the \citeauthor{bower06} model and 
the dashed line shows the \citeauthor{baugh05} model. The compilation of observed counts 
was kindly supplied by Nigel Metcalfe and can be downloaded from 
http://star-www.dur.ac.uk/$\sim$nm/pubhtml/counts/kdata.txt\,.
}
\label{fig:kdata}
\end{figure}

The local K-band luminosity function is one of the datasets used to fix the 
values of the parameters which define each semi-analytical model. A comparison 
of the model predictions to these data is given in \citet{baugh05} and \citet{bower06}. \citeauthor{bower06} also show 
how the \citeauthor{baugh05} and \citeauthor{bower06} models compare against observational estimates of the 
rest-frame K-band luminosity function to z=1.5. In the \citeauthor{bower06} model, there is 
remarkably little evolution at the bright end with redshift, in very good 
agreement with the observations. The \citeauthor{baugh05} model does predict a modest but 
significant degree of evolution at the bright end. 

Before considering the surface density of EROs, it is instructive to first look at 
the overall K-band number counts, which are sensitive to the evolution of 
the {\it observer} frame K-band luminosity function. 
The motivation for this is the following. A deficiency in the 
predicted number counts of EROs could be explained in part by the failure of 
a model to match the total galaxy counts. Fig.~\ref{fig:kdata} shows\footnote{The observed K-band counts are from the following sources, with symbols indicated in the Fig.~\ref{fig:kdata} legend: \citet{vaisanen00,d95,mcc00,mob86,gardner93,gardner96,glaze95,mcl95,soifer94,huang97,huang01,moustakas97,saracco97,saracco01,metcalfe01,ber98,sz98,koch01,mamon98,martini01,vandame01,kummel01,mai01,kong06}} the 
differential K-band number counts, after dividing by a power law fit to  
the observed counts to expand the useful dynamic range on the y-axis. What 
is plotted is effectively the deviation of the counts from the power law that best fits the compiled observational data. There is a considerable spread in the observed counts at bright 
magnitudes ($K<15$), which could be a sign that surveys in particular directions 
are affected by a local hole in the galaxy distribution \citep[e.g.][]{frith06}.  
The \citeauthor{bower06} model is higher than many of the observed counts at these magnitudes, 
but does agree well with some of the datasets. The \citeauthor{baugh05} model on the other hand 
underpredicts the bright counts by a factor of three. This discrepancy can be 
traced to a mismatch between the predicted and observed shape of the K-band luminosity 
function around $L_*$. The observed number counts are most sensitive to the form of 
the luminosity function close to $L_*$, whereas when assessing a plot of a 
luminosity function, the eye is naturally drawn to the faint and bright ends. 
The scatter between the various observational estimates of the counts is 
still almost a factor of two at faint magnitudes. At $K\sim 19$, the \citeauthor{baugh05} counts are a factor of four lower than the prediction of 
the \citeauthor{bower06} model. The \citeauthor{baugh05} and \citeauthor{bower06} model 
predictions bracket the observational data at $K\sim 19$ and converge at $K\sim 23$.

\subsection{The abundance of EROs}\label{sec:N_EROs}

\begin{figure}
{\epsfxsize=8.5truecm
\epsfbox[33 20 509 488]{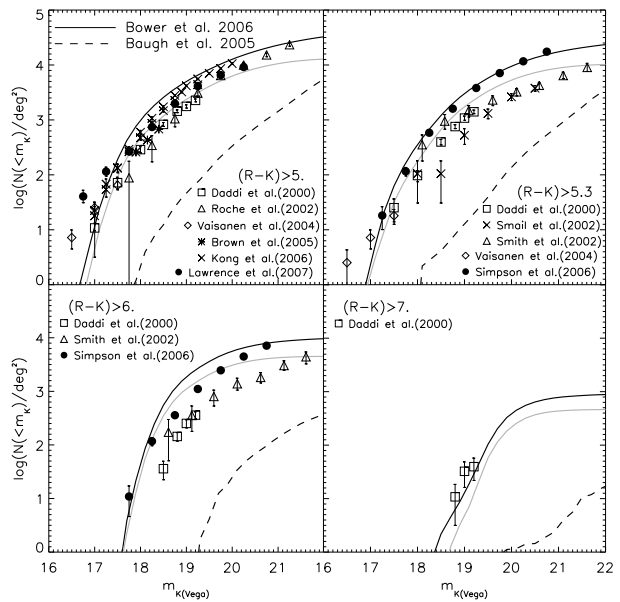}}
\caption
{ The cumulative K-band number counts for EROs selected by their $(R-K)$
  colour: $(R-K)>5$ -- top left panel, $(R-K)>5.3$ -- top
  right panel, $(R-K)>6$ -- bottom left panel and $(R-K)>7$ in the
  bottom right panel. The solid black lines corresponds to the predictions
  from \citeauthor{bower06} and the dashed ones to the \citeauthor{baugh05} model. 
  The grey lines show the number counts of EROs in the \citeauthor{bower06} model 
  that are also central galaxies. The source of the observational 
  data is given in the legend of each graph and the points can be found at: 
  http://segre.ieec.uab.es/violeta/rawEROSdata.txt. The errors shown are Poisson.
}
\label{fig:erodndm}
\end{figure}

As we discussed in the Introduction, reproducing the abundance of EROs 
has previously eluded many hierarchical models, but is a necessary 
requirement for any model which aims to explain the nature of these 
objects. In Fig.~\ref{fig:erodndm} we compare the observed counts\footnote{The observed ERO  
counts are from the following sources, with symbols indicated in Fig.~\ref{fig:erodndm} legends: \citet{daddi00,roche02,smith02,smail02,vaisanen04,brown05,simpson06,kong06,lawrence07}}
with the predicted counts for samples of EROs defined by different colour cuts, 
$(R-K)>5,5.3,6$ and $7$. The dispersion between observational estimates 
is due in part to the use of different filters, the application of different 
apertures to determine colours and sample variance due to the small fields used 
\citep{simpson06}. The predictions of the \citeauthor{bower06} model reproduce the 
observed abundance of EROs impressively well, for all the colour 
cuts shown, which is remarkable as none of the model parameters have been 
tuned to achieve this level of agreement. The observational data for EROs 
defined by $(R-K)>5$ show the least dispersion. These results are 
matched well by the \citeauthor{bower06} model, except around $K=18$ where the model 
slightly overpredicts the number counts by a factor of $\sim 2$. 
\citeauthor{bower06} overestimates the number counts of EROs with $(R-K)>6$, 
particularly around $K=19$ where the discrepancy is close to a factor of five. 

The excess of EROs in the Bower et~al. model could result from the star 
formation in massive galaxies being quenched too efficiently at high redshift 
by the AGN heating of the hot halo, which cuts off the supply of cold gas and hence 
leads to old stellar populations. An alternative explanation 
could lie in the treatment of satellite galaxies in the model. Typically in 
semi-analytical models, when a halo merges with another halo, the hot gas is assumed 
to be stripped completely from the smaller halo. The largest galaxy in the newly 
formed halo is called the central galaxy, and any gas which subsequently cools is 
directed onto this galaxy. Star formation in the satellite ceases when its reservoir of 
cold gas has been exhausted; this is sometimes referred to as 
``strangulation''. Motivated by the results of gas dynamics simulations of 
the efficiency of hot gas stripping by ram pressure from McCarthy 
et~al. (2008), Font et~al. (2008) presented a revised version of {\tt GALFORM}
in which satellites retain some fraction of their hot gas haloes, depending 
on their orbits. Font et~al. found that this new model changes the fraction of 
satellites with red colours and gives a better match to the colour distribution 
of satellites in galaxy groups as determined from the Sloan survey by Weinnmann et~al. 
(2006). However, this extension of the gas cooling model is unlikely to have a significant  
impact on the predicted counts of EROs, since the galaxies whose colours are affected are 
primarily fainter than $L_*$. We show the contribution of central galaxies to the 
counts of EROs in Fig.~\ref{fig:erodndm}. At faint magnitudes, satellites account for 
a substantial fraction of the ERO population, and the predictions of the Bower et~al. 
model are in better agreement with the observed counts if satellites are excluded 
altogether. The Font et~al. paper was in the final stages of being refereed when our 
paper was being prepared for submission; the number counts of EROs in the Font et~al. 
model will be discussed in the other papers in this series.

The agreement is less impressive for the \citeauthor{baugh05} model. 
Fig.~\ref{fig:erodndm} shows that this model typically underestimates 
the number counts of EROs by more than an order 
of magnitude. Appealing to scatter resulting from the small size of the fields 
surveyed to measure the ERO counts is overly optimistic. The errors shown in 
Fig.~\ref{fig:erodndm} are Poisson counting errors. EROs are believed to be strongly 
clustered (see Paper II) and so the Poisson error is a lower limit on 
the error, to which sampling variance in the spatial distribution of EROs should be 
added. However, it has been argued that this is unlikely to account for more 
than a factor of two scatter in number density \citep{somerville04}. Another possibility 
to improve the \citeauthor{baugh05} predictions would be to force the model to agree 
with the overall $K-$band counts. In the previous section, we pointed out that 
at $K\sim 19$ the \citeauthor{baugh05} model underpredicts the total counts by a 
factor of $\sim 4$. One could imagine applying an equal correction to the 
predicted luminosities in all bands (to preserve galaxy colours), in order to force 
the \citeauthor{baugh05} K-band counts to agree with the observations. 
Upon doing so there would still be a factor of 2.5 to 3 discrepancy 
with the observed ERO counts. 

Some fraction of EROs are likely to be heavily dust extincted, a point we return 
to in Section \ref{sec:mix}. It is therefore important to make a realistic 
calculation of the degree to which galaxy colours are extincted in order to be able 
to make robust predictions for the counts of EROs. As we shall see in the next subsection, 
the typical redshift of EROs with $(R-K)>5$ is $z\approx 1$. The R-band samples the rest 
frame near ultra-violet ($\sim 3300 \AA{}$) for a galaxy at this redshift. In our standard 
calculations we use a refinement of the dust extinction model introduced by \citet{cole00}, 
as described in Section 2, in which the dust is assumed to be in two components, diffuse 
dust and molecular clouds. We have experimented with changing the fraction of 
dust contained in molecular clouds (the default choice is 25\%) and do not find a 
significant difference in the predicted counts of EROs in the \citeauthor{baugh05} model. 
We also repeated the calculation using the spectro-photometric code {\tt GRASIL} 
\citep{granato00,lacey08}. Again, we find a similar amount of extinction to that obtained 
with the less expensive calculation carried out with {\tt GALFORM} alone.

We shall see in the next section that the \citeauthor{baugh05} model predicts very few dusty starburst EROs. 
This can also be connected to the lack of a break in the predicted shape of the ERO 
counts in \citeauthor{baugh05}. The observations of both \citet{kong06} and \citet{smith02} show evidence for  
the presence of a break in the slope of number counts at $K\in[18,19]$ for EROs 
with $(R-K)>5$ and $K\in[19,20]$ for EROs with $(R-K)>5.3,6$. This has led 
to speculation that there may be a change in the nature of EROs around this magnitude, with  
a transition from the dominance of old galaxies at low redshifts and to dusty starbursts 
at high redshift. No such feature is apparent in the \citeauthor{baugh05} model, except for EROs with 
$(R-K)>6$. To investigate if the lack of a break (and the overall deficit of ERO counts) 
is due to the choice of burst timescale adopted in \citeauthor{baugh05}, we reran the model with a burst 
time scale set to twice the bulge dynamical time, with a minimum allowed value of $1$ Myr. 
The original parameter values adopted in \citeauthor{baugh05} were 50 times the dynamical time with a 
minimum timescale of 200 Myr. Changing to a shorter burst duration had little impact 
on the predicted number of EROs. 

The disagreement between the \citeauthor{baugh05} model and observations suggests that the  
superwind feedback used in this model could be too efficient for bright galaxies at $z>1$, 
delaying their formation, and that bursts of star formation maybe should also 
be allowed to be triggered by other events besides mergers (such as the dynamical 
instabilities of disks), in order to try to account for the larger observed number 
of red, dusty star-forming galaxies.

\subsection{The redshift distribution of EROs}\label{sec:nz}

\begin{figure}
{\epsfxsize=8.5truecm
\epsfbox[0 33 235 387]{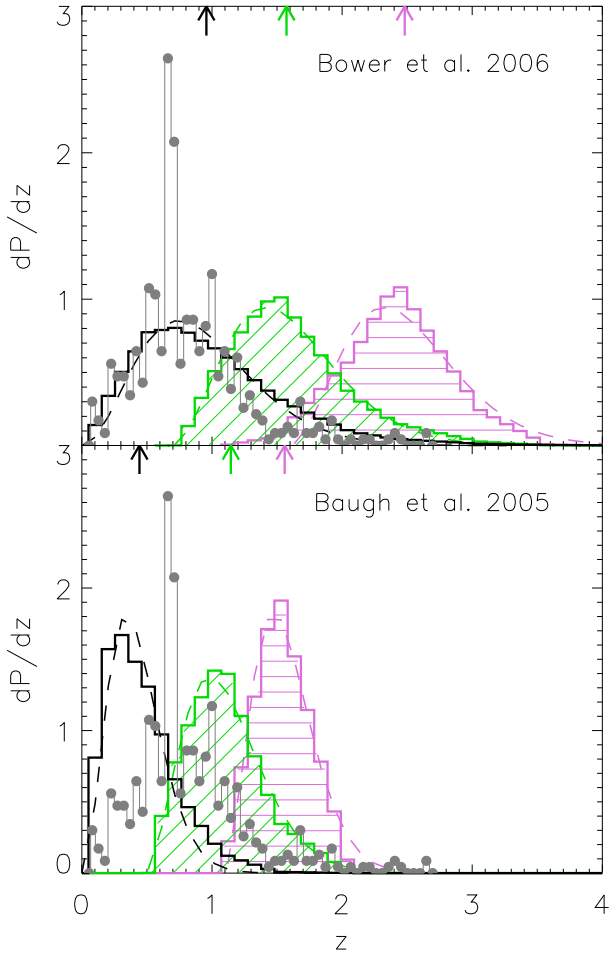}}
\caption
{
The redshift distribution of EROs with $K\leq 20$, defined by $(R-K)$ colour redder
 than $5$ (histogram filled with sloping parallel lines) and $7$ (histogram filled with parallel lines). The black histogram represents 
the redshift distribution for all galaxies with $K\leq 20$. Histograms are normalized to give unit area under each distribution. The arrows on the upper x-axis 
show the median of the distribution plotted with the same colour. 
Observational data from \citet{cimatti02b}
for galaxies with $K\leq 20$  are shown as grey connected filled circles. 
The upper panel shows the predictions from \citeauthor{bower06} and the lower one 
from the \citeauthor{baugh05} model. The dashed curves show parametric fits to the 
model redshift distributions (see Eq. \ref{eq:nz} and Table \ref{tab:dndz}). 
}
\label{fig:nz1}
\end{figure}

\begin{figure}
{\epsfxsize=8.5truecm
\epsfbox[0 33 235 387]{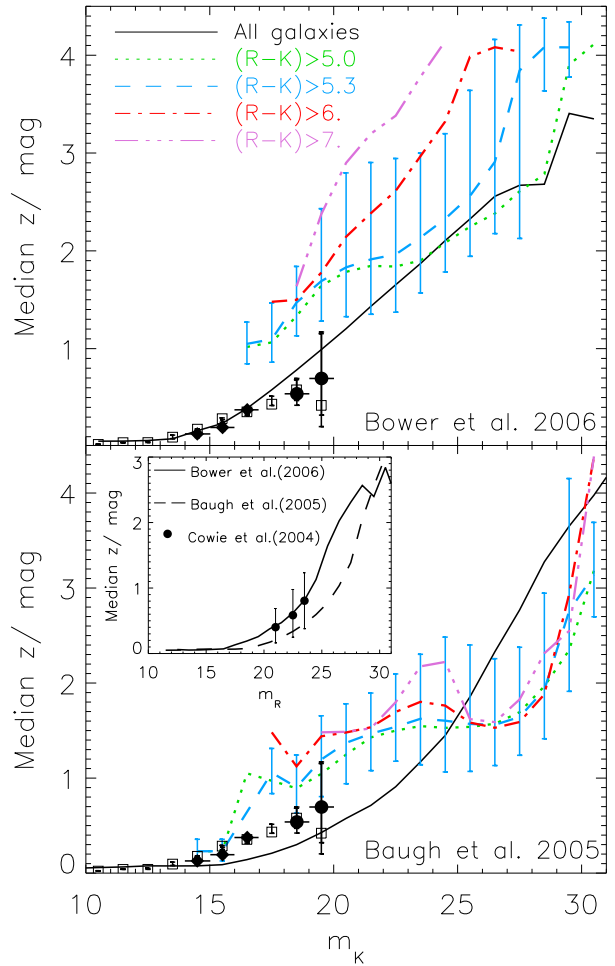}}
\caption
{The median redshift as a function of K band magnitude for all galaxies (solid line) and
for EROs with $(R-K)$ colour redder than  $5$ (dotted line), $5.3$ (dashed line), $6$ (dash-dotted line) and 
$7$ (dash-triple dotted line). For clarity, quartiles are only shown for EROs with $(R-K)>5.3$. 
The upper panel shows predictions from \citeauthor{bower06} and the lower one 
from \citeauthor{baugh05}. For comparison we plot the median redshift for all galaxies 
per magnitude bin in the $K$-band observed by \citet{songaila90} (squares), 
\citet{glazebrook95} (diamonds) and \citet{cowie96} (circles). The inset shows the 
median redshift per magnitude bin in the $R$-band for all galaxies predicted 
by \citeauthor{bower06} (solid line) and \citeauthor{baugh05} (dashed line), and 
the observational data from \citet{cowie04} (filled circles).
}
\label{fig:zmedian}
\end{figure}
 
The colour criteria that classify a galaxy as an ERO are designed 
to select objects at $z\geq 1$ based on combining simple star formation 
histories with stellar population synthesis models \citep[e.g.][]{thompson99}. 
However, the redshift selection can be different when dealing with 
hierarchical galaxy formation models, due to the 
rich variety of star formation histories that can be generated \citep[see for example][]{baugh06}. In this section we present the redshift distributions of EROs predicted 
by the \citeauthor{bower06} and \citeauthor{baugh05} models. 

We start by testing the model predictions against observational data for K-selected 
galaxies, since EROs are a subsample of these. The predicted differential redshift 
distributions of galaxies brighter than $K= 20$ are shown in Fig.~\ref{fig:nz1}, 
and are compared with the observational data from the K20 survey \citep{cimatti02b}. 
This survey covers an area of 52 ${\rm arcmin}^{2}$ and contains 480 galaxies with $K\leq 20$. 
The spectroscopic completeness is $87\%$; most of the incompleteness is for galaxies 
at $z>1$, for which it was particularly hard to extract redshifts, unless strong H$\alpha$ 
was present in emission. Where a spectroscopic redshift was not available, a photometric 
redshift was estimated using multiband photometry. Observations by \citet{cirasuolo07} 
find a similar redshift distribution to that of the K20 survey, using a larger sample 
of galaxies, $\sim 2200$ in 0.6 $deg^{2}$, but with only $545$ spectroscopic detections and 
so a much heavier reliance on photometric redshifts. The spike in the K20 redshift distribution at 
$z\sim 0.7$ is due to the presence of two clusters. Ignoring this spike, it can be seen 
that \citeauthor{bower06} makes an accurate prediction of the redshift distribution 
of K-selected galaxies out to $z\sim 1.3$. At higher redshifts, the \citeauthor{bower06} 
model overpredicts the number of objects. This is reinforced by the fact that the median 
redshift predicted by \citeauthor{bower06} model, $z_{\rm med}=0.96$, is comparable to but 
slightly higher than that observed by \citet{cimatti02b}, $z_{\rm med}=0.805$ (excluding the 
two clusters; $z_{\rm med}=0.665$ otherwise). Such good agreement is in contrast with 
the \citeauthor{baugh05} model which predicts a narrower redshift distribution around a 
lower redshift value, $z_{\rm med}=0.42$. Fig.~\ref{fig:nz1} shows that the observation of 
significant numbers of galaxies with $z>1$ favours the \citeauthor{bower06} model; in the \citeauthor{baugh05} model 
it is much less likely that one would find galaxies at such redshifts. 

\begin{table*}
 \centering
 \begin{minipage}{140mm}
 \caption{
{\small Best fits to the redshift distributions predicted by 
the \citeauthor{bower06} and \citeauthor{baugh05} models, as described 
in \S\ref{sec:nz}. The first column gives the $K$-band magnitude limit and the 
second gives the colour selection applied. Columns 3, 4, 5, 6 give the best fitting 
parameters and the value of the summed squared residuals under $\chi^{2}$ for the \citeauthor{bower06} model; columns 7 to 10 give the 
equivalent quantities for the \citeauthor{baugh05} model. 
} }
\begin{tabular}{ccccccccccc}
\hline
\multicolumn{2}{c}{ }  &
\multicolumn{4}{c}{Bower et al.} &
\multicolumn{1}{c}{ }  &
\multicolumn{4}{c}{Baugh et al.} \\ 
\cline{3-6}
\cline{8-11}
$K\leq$ & $(R-K)>$ & $A$ & $z_{c}$ & $\alpha$ &$\chi ^{2}$ & & $A$ & $z_{c}$ & $\alpha$ & $\chi ^{2}$ \\
\hline
20. & none & 5.825  &  0.616 &    -   &  0.100 & & 58.111 &  0.284 &     -    & 0.359 \\
    & 5.0 & 6.681  &  0.604 & 0.683 &  0.032 & & 19.846 &  0.424 &  0.474  & 0.067 \\
    & 5.3 & 11.222 &  0.501 & 0.930 &  0.074 & & 10.996 &  0.519 &  0.493  & 0.059 \\
    & 6.0 & 10.352 &  0.518 & 0.968 &  0.085 & & 29.846 &  0.367 &  0.996  & 0.297 \\
    & 7.0 & 6.608  &  0.608 & 1.588 &  0.199 & & 44.581 &  0.325 &  1.083  & 0.131 \\
18. & 5.3 & 55.775 &  0.297 & 0.723 &  0.062 & & 73.358 &  0.265 &  0.737  & 0.333 \\
22. & 5.3 & 6.014  &  0.616 & 0.890 &  0.086 & & 14.774 &  0.464 &  0.859  & 0.209 \\
\hline
\end{tabular}\label{tab:dndz}
\end{minipage}
\end{table*}

For further comparison, we plot in Fig.~\ref{fig:zmedian} the observed median redshift 
distributions for both the K \citep{songaila90,glazebrook95,cowie96} and R bands 
\citep{cowie04} and compare with the model predictions. The \citeauthor{bower06} 
model is consistent with the observations in both bands. The \citeauthor{baugh05} 
model matches the data better in the $R$-band than it does in the $K$-band. 
The \citeauthor{baugh05} model underestimates the median redshift for K-selected samples faintwards 
of $K\sim 16$. 

Both Fig.~\ref{fig:nz1} and Fig.~\ref{fig:zmedian} show predictions for samples 
of EROs defined by different colour cuts. The most common way of selecting EROs is by 
imposing $(R-K)>5.3$. Nevertheless different definitions have been used in observational 
studies which results in different properties \citep[see][]{mccarthy04}. 
The \citeauthor{bower06} (\citeauthor{baugh05}) model predicts the median redshifts to be 
$z_{\rm med}=1.57 (1.12), 1.77 (1.52)$ and $2.48 (1.53)$ for galaxies with $K\leq 20$ redder than $(R-K)=5,6$ 
and $7$, respectively. As expected from stellar population synthesis 
models, the median redshift of EROs is $z_{\rm med} \sim 1$ or higher, independent 
of the colour cut and model; this also explains why there are few EROs at very bright K-band 
magnitudes.  
\citet{cimatti03} measured a median redshift of $z_{\rm med}\sim 1.27$ for a sample of EROs 
with $(R-K)>5$ and $K\leq 20$, which was only $62\%$ complete spectroscopically. 
Other surveys using a mixture of spectroscopic and photometric redshift determinations 
find similar results: for EROs with $(R-K)>5$, \citet{moustakas04}, with a limiting 
magnitude of $K=20$, measured a median redshift of $z_{\rm med}\sim 1.20$, whereas 
\citet{brown05}, with a brighter cut of $K<18.4$, found $z_{\rm med} \sim 1.18$.
As expected from the above discussion, these values are somewhat higher than the 
predictions of the \citeauthor{baugh05} model and lower than the \citeauthor{bower06} 
predictions.

Within the UKIDSS survey data, \citet{simpson06} found a clear tendency for redder EROs 
to be at higher redshifts. In Fig.~\ref{fig:nz1} the same tendency is predicted 
by the \citeauthor{bower06} model. This trend is also present in the \citeauthor{baugh05} 
model but is less pronounced than in the \citeauthor{bower06} model. 
According to Fig.~\ref{fig:nz1} and Fig.~\ref{fig:zmedian}, the \citeauthor{baugh05} 
model predicts few galaxies beyond $z \sim 1.5$ in simple K-selected samples. However, 
on applying a colour cut, the median redshift increases sharply: for galaxies 
with $K<20$ and $(R-K)>6$, the median redshift is $z \sim 1.5$. Thus, even in the 
\citeauthor{baugh05} model, which on the whole tends to predict shallower redshift distributions 
than are observed, redder ERO samples probe higher redshifts ($z > 1$).   

The shape of the redshift distribution can be parameterized using the 
simple parametric form introduced 
by \citet{efsta91}:
\begin{equation}\label{eq:nz}
\frac{{\rm d}N}{{\rm d}z} = A z'^{2} \exp \left[ - \left( \frac{z'}{z_{c}} \right) ^\beta \right],  \quad \quad {\rm for \,\,} z' > 0,
\end{equation}
where $A$ is a normalization parameter in units such that ${\rm d}N/{\rm d}z$ gives 
the number of galaxies per square degree per unit redshift, 
$\beta$ controls the shape of the distribution 
and $z_{c}$ is related to the median redshift of the distribution as a function of 
$\beta$. EROs are not found at low redshifts, therefore we introduce an offset to 
the distribution, $z'=z-\alpha$. When fitting the redshift distribution of K-selected 
galaxies we do not need this offset, since these galaxies are expected to populate 
the low redshift range.

The best fit parameters for the redshift distributions predicted by the 
\citeauthor{bower06} and \citeauthor{baugh05} models are listed in Table \ref{tab:dndz}. 
\citet{baugh93} found that $\beta=3/2$ gave a good description of the 
shape of the redshift distribution for B-selected samples. We find that the same 
value of $\beta$ also gives a good match to the shape of the samples considered 
here, so we hold $\beta=3/2$ fixed. As can be seen in both Table \ref{tab:dndz} 
and Fig.~\ref{fig:nz1}, the empirical model described by Eq.~\ref{eq:nz}, accurately 
reproduces the model predictions. The median redshift $z_{\rm med}$ is related to $z_{c}$ 
for $\beta=3/2$ as follows \citep{baugh93}: 
\begin{equation}\label{eq:zm}
z_{\rm med} = 1.412z_{c} + \alpha.
\end{equation}
Extracting the median redshift from Table \ref{tab:dndz}, we again find the tendency 
for redder galaxies to be at higher redshifts. 

The last two rows in Table~\ref{tab:dndz} list the best fit parameters for the redshift 
distribution of EROs redder than $(R-K)=5.3$ with magnitudes limits different from $K=20$. 
The median redshift is readily obtained directly from the fit; for the \citeauthor{bower06}
(\citeauthor{baugh05}) model the median derived in this way is $z_{\rm med}=1.13(1.11),1.65(1.23)$ and $1.85(1.53)$ 
for EROs with $(R-K)>5.3$ and brighter than $K=18,20$ and 22, respectively. 
Thus from this it is clear that with progressively fainter limits we observe more 
distant galaxies.

By its definition, $\alpha$ gives an estimation of the redshift at which EROs start 
to appear. It can be seen from Table \ref{tab:dndz}, that for $K\leq 20$, EROs defined 
with redder cuts start to appear at higher redshifts. However, this tendency is less 
clear for fainter magnitude limits. This could be due to the fact that samples defined 
by $K>20$ already probe higher redshifts, so there is less variation on changing the 
colour cut. \citet{bergstrom04} studied in detail at what redshift galaxies can be 
classified as EROs, based on the spectral evolution synthesis model {\sc PEGASE2}. 
They found out that galaxies with star formation histories matching those inferred 
for ellipticals will show colours as red as $(R-K)=5$ for redshifts $z\geq 0.6$, 
depending on the metallicity assumed. These authors also found that starburst galaxies 
could be detected as EROs, i.e. with $(R-K)>5$, at redshifts around $z=5$ for an 
extinction of $E(B-V)\leq 1$, or at lower redshifts for stronger extinctions, 
on assuming a \citet{calzetti00} extinction law. 
The redshift that \citet{bergstrom04} proposed for the emergence of passively 
evolving EROs is similar to the prediction for all EROs by the \citeauthor{bower06} 
model, though is a bit higher than suggested by the \citeauthor{baugh05} model. 

\begin{figure}
{\epsfxsize=8.5truecm
\epsfbox[10 8 237 232]{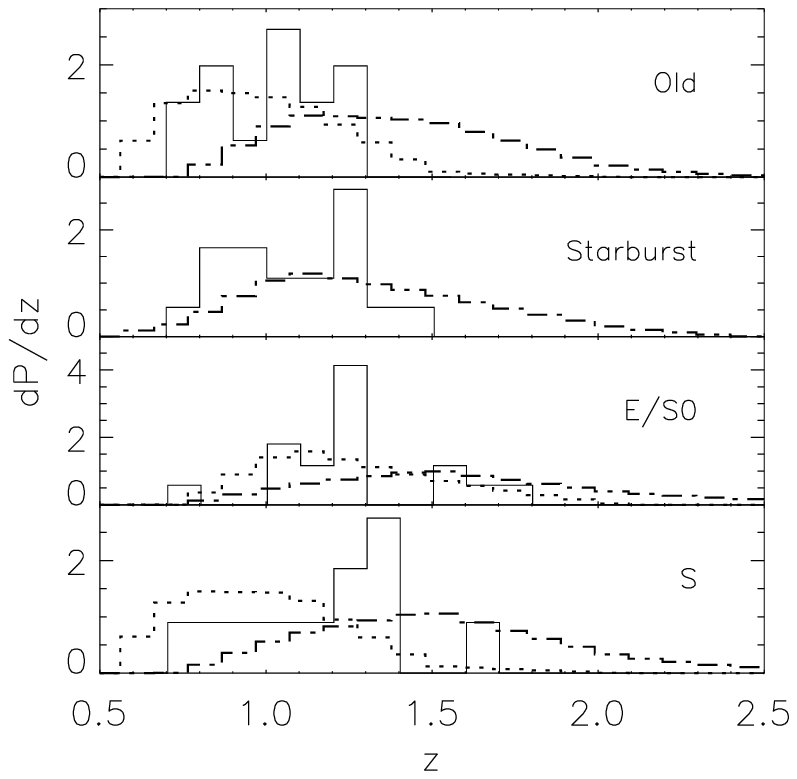}}
\caption
{
The redshift distribution of EROs with $(R-K)>5.0$, predicted by \citeauthor{bower06} 
(dash-dotted line) and \citeauthor{baugh05} (dotted line), compared with the 
observational data from \citet{cimatti02a} and \citet{cimatti03} (solid line). 
The magnitude limits for the EROs are $K=19.2$ for the two upper panels and $20$ 
for the two lower panels respectively. From top to bottom the redshift distributions are plotted 
for quiescent or old galaxies, galaxies that are experiencing a burst of star formation (note that 
there is no dotted line in this panel, as the Baugh et~al. model does not predict starburst EROs), 
spheroidal galaxies and spiral ones (see text for further details on the classification). Histograms are normalized to give unit area under each distribution.
}
\label{fig:nz_types}
\end{figure}

Since the work of \citet{bergstrom04} is based on combining stellar population synthesis 
models with simple star formation histories, it is important to compare the semi-analytical 
model predictions directly with observations. Fig.~\ref{fig:nz_types} shows the predictions 
of the \citeauthor{bower06} and \citeauthor{baugh05} models for the redshift distribution 
of EROs with $(R-K)>5$ classified according to either their star formation activity or 
morphological type. 
In this plot, the distinction between the \textit{old} and \textit{starburst} populations 
depends on whether or not the galaxy is undergoing or recently experienced an episode 
of star formation triggered by a galaxy merger or by a dynamically unstable disk. A model ERO is 
classified as a \textit{starburst} if it is undergoing such a starburst, or if one 
happened within the past 1 Gyr. Note that we investigate an alternative definition of activity in Section 4.2 based on the 
specific star formation rate, and find that, with a suitable choice of specific star 
formation rate, we obtain similar numbers of ``active" and ``passive" galaxies as we 
do when the activity is considered to have been triggered by some event, as is the case in this section.
For morphological classification, we use the bulge-to-total 
luminosity ratio in the observer-frame K-band, $B/T$. Further details of the definitions of 
the various classes of galaxies will be given in section \ref{sec:mix}.

It can be seen in Fig.~\ref{fig:nz_types} that both the \citeauthor{bower06} and 
\citeauthor{baugh05} models predict that quiescent galaxies appear as EROs at 
redshifts $z\geq 0.7$, in agreement with the expectations of \citet{bergstrom04}. 
Nevertheless, both models show starburst galaxies with ERO colours at redshifts 
lower than $z=5$, though none of these galaxies are predicted 
by the \citeauthor{bower06} model to have $E(B-V)> 1$ (see next section), as was 
assumed (albeit for a different extinction model) by \citet{bergstrom04}. Fig.~\ref{fig:nz_types} includes \citet{cimatti02a} observational data. \citet{cimatti02a} classify galaxies as old if they have a prominent $4000 \AA{}$ break 
and no emission lines, and as starbursts if they do show emission lines. 
The \citeauthor{baugh05} redshift distribution of old EROs matches the observed one. 
However, it is notable that this model predicts no dusty starbursts with the colours of 
EROs. In the \citeauthor{baugh05} model, there are sufficient dusty starbursts to account for the 
sources detected by SCUBA, mostly at somewhat higher redshifts than relevant here for 
EROs. The \citeauthor{bower06} model overpredicts the high redshift tail of the starburst distribution, however the incompletness of observations at these redshifts could account for this mismatch.

Fig.~\ref{fig:nz_types} also shows the predicted redshift distribution for different morphological 
types compared to the observational 
data from \citet{cimatti03}. Their classification is based upon visual inspection of 
images in the observer-frame B, V, i and z bands. It can be seen that \citeauthor{baugh05} model 
predicts a redshift distribution for spheroidal EROs which is quite close to the 
observed one, while the \citeauthor{bower06} model overpredicts, again, the high redshift tail.
In the case of disk-dominated galaxies the mismatch for both models increases, 
with the \citeauthor{baugh05} model predicting too few high redshift galaxies 
and \citeauthor{bower06} too many.

In summary, the \citeauthor{baugh05} model tends to predict shallower redshift 
distributions than the observations for all EROs, while \citeauthor{bower06} is in better 
agreement with the data. However, the \citeauthor{baugh05} model better matches the 
observed redshift distribution of both old and spheroidal EROs. It seems that 
the \citeauthor{baugh05} model underpredicts the number of disk dominated galaxies, 
particularly at redshifts higher than $z\sim 1$. For different types of EROs 
the \citeauthor{bower06} model consistently overpredicts the number of galaxies 
in the high redshift tail. This is perhaps an indication that the star formation is 
quenched too early in massive galaxies in this model. However, it should be borne 
in mind that the observations we are comparing to are affected by cosmic variance and 
by incompleteness at high redshift.

\section{The nature of EROs}\label{sec:mix}

In the previous section we presented the predictions of the \citeauthor{baugh05} 
and \citeauthor{bower06} models for the number counts and redshift distribution 
of EROs. The \citeauthor{bower06} model gave a better match to the observations 
in both instances, and for this reason we restrict our investigation of the 
predicted properties of EROs to this model. In addition to underpredicting the 
abundance of EROs by a large factor, the \citeauthor{baugh05} model does not 
predict any dusty starburst EROs. Therefore the properties of EROs in this model are 
of limited interest, since we do not know how they compare with the properties of 
the ``missing'' EROs, which could be substantially different.

In this section, we investigate whether or not EROs form one or more distinct 
populations of galaxies, or if there is a continuous range of intrinsic properties, 
marking a gradual transformation of active, massive galaxies to passive objects, 
as suggested by \citet{conselice08}. Observationally, EROs exhibit a mixture of 
morphologies and spectral properties \citep[e.g.][]{smail02}. At the epoch when 
EROs are detected, $z\geq 1$, the star formation rate per unit volume is much higher 
than it is today and the range of galaxies which are undergoing significant additions 
to their mass is much broader than is inferred locally \citep[e.g.][]{abraham07}. 
Thus, the connection between a galaxy's morphology and its spectral energy distribution 
cannot be assumed to be the same as it is for nearby galaxies. In fact, \citet{smail02} 
pointed out that the visual classification of EROs did not match their spectral energy 
distributions (SED). Consequently, in this section we study the nature of EROs by analyzing 
separately their morphology and their level of star formation activity, the primary agent 
which shapes a galaxy's SED. 

\subsection{The morphology of EROs}

\begin{figure}
{\epsfxsize=8.5truecm
\epsfbox[49 33 512 488]{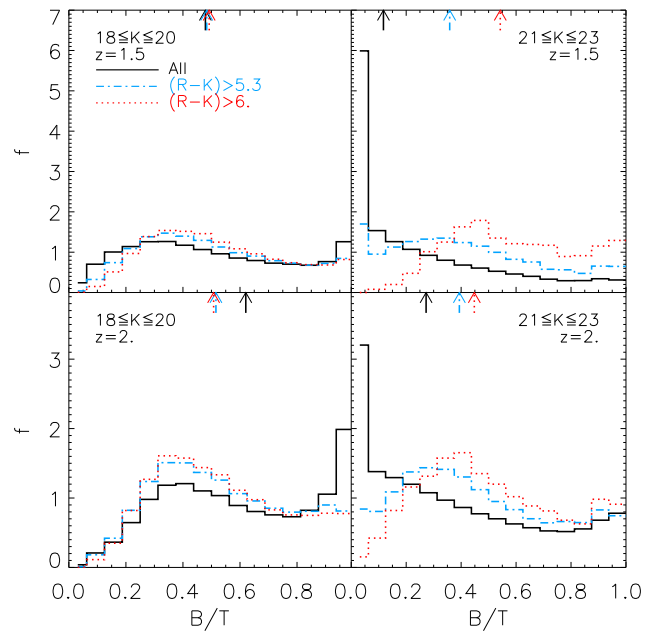}}
\caption
{
The predicted distribution of galaxy bulge to total luminosity ratio, in the observer frame K-band, in the \citeauthor{bower06} model. 
The top row shows $z=1.5$ and the bottom shows $z=2$. The panels in the left-hand column correspond 
to an apparent magnitude range of $18 \leq K \leq 20$ and the right hand panels show $21 \leq K \leq 23$. $K$-selected galaxies distribution appear as a solid line. EROs distributions are represented by the dotted, $(R-K)>5.3$, and dash-dotted histograms, $(R-K)>6$. The arrows show the median bulge to total ratio for the histogram plotted with the same line style.
}
\label{fig:bt}
\end{figure}

Semi-analytical models track the amount of light in the spheroid and disk components 
of model galaxies \citep[e.g.][]{baugh96b}. A simple proxy for morphological 
classification is therefore the bulge to total luminosity ratio, $B/T$.
Here, we use this ratio measured in the observer frame K-band to apply a morphological 
classification to the model galaxies. Different observational studies have used images 
in different bands to make morphological classifications, so our choice is necessarily a 
compromise and we are more interested in the trends with $B/T$ ratio than with absolute 
values. 
There is a large scatter between the bulge to total luminosity ratio and other 
indicators of morphology such as the subjective Hubble T-type \citep[][]{simien86}. 
Despite this, we follow the results of Simien \& de Vaucouleurs and adopt a 
boundary between early-type (bulge dominated) galaxies and late-type 
(disk dominated) galaxies at $B/T =0.4$. Our results are insensitive to the 
precise choice of this value. We note that one potential flaw in this scheme 
is that irregular and interacting 
galaxies are not naturally accommodated into the $B/T$ classification whereas, observationally, 
they become a more important population with increasing redshift, particularly around 
the typical redshift of ERO samples \citep[e.g.][]{abraham07}. Ultimately, the broad brush 
nature of our morphological classification will limit how closely we can compare the model 
predictions to observational results. 

Fig.~\ref{fig:bt} shows the predicted distribution of $B/T$ in the \citeauthor{bower06} model for $z=1.5$ and $z=2$ 
in two different apparent magnitude ranges. This plot shows that it is more likely to 
find K-selected galaxies with late-type $B/T$ ratios in the fainter magnitude range than it is 
in the brighter magnitude bin. The same tendency is found for EROs, although it is weaker since the 
variation from one magnitude range to the other is smaller in this case than it is when 
simply K-selecting galaxies without a colour cut. Averaging over the redshift interval 
$1\leq z\leq 2.4$, the predicted median $B/T$ values for EROs with $(R-K)>5.3\, (6)$ 
shifts from $0.42\,(0.48)$ in the faint bin to $0.51\,(0.53)$ in the bright one. The 
same tendency is observed by \citet{moustakas04}, when considering as late-type EROs 
those not classified as early-type. Note, however, that these authors classified 
some interacting EROs as having early-type morphologies, which makes a direct comparison with 
our classification based on $B/T$ values difficult.  

\begin{table}
\caption{
{\small 
The predicted morphological mix of galaxies in the \citeauthor{bower06} model, 
expressed as a percentage of \textit{early-type:late-type}, for galaxies 
with $K\leq 23$ at $z=1,1.5,2$ and $2.5$, including EROs redder than 
$(R-K)=5,5.3,6$ and $7$. We use the bulge to total luminosity in the 
observer frame $K$-band as a proxy for morphology, with $B/T=0.4$ setting the 
boundary between early and late types.}  
}
\begin{center}
\begin{tabular}{ccccc}
\hline
 &  \multicolumn{4}{c}{$z$}  \\
\cline{2-5}
$(R-K)$ & 1.0 & 1.5 & 2.0 & 2.5\\
\hline
All & 28:72  & 35:65 & 45:55  & 53:47 \\
5.0 & 63:37  & 55:45 & 59:41  & 65:35 \\
5.3 & 59:41  & 59:41 & 60:40  & 66:34 \\
6.0 &   -    & 67:33 & 66:34  & 69:31 \\
7.0 &   -    &   -   & 67:33  & 70:30 \\
 \hline
\end{tabular}\label{tab:morfo}
\end{center}
\end{table}

Table~\ref{tab:morfo} gives the predicted morphological mix of K-selected galaxies with $K\leq23$, 
and EROs at redshifts $z=1,1.5,2$ and $2.5$. A slight increase is apparent in the percentage 
of early-type K-selected galaxies with increasing redshift. This tendency is due to the fact that for a 
fixed apparent magnitude range, when we increase the redshift we are looking at intrinsically 
brighter galaxies, which, as we have seen in Fig.~\ref{fig:bt}, are predominantly early-type.
The same behaviour is seen for EROs. Independent of the colour cut used to define EROs, 
for $z>1$ there is a slight increase with redshift in the percentage of EROs with early-type 
$B/T$ ratios. It should be noted that due to the form of the redshift distribution of faint 
EROs with $(R-K)>6$, the statistics at $z=1.5$ are actually poorer than they are at $z=2$. 
For $(R-K)>6$ EROs, the increase in the proportion of galaxies that are early-types is clearer 
on comparing the change in $B/T$ seen when moving from $z=2$ ($B/T_{\rm median}=0.47$), to $z=2.4$ 
($B/T_{\rm median}=0.49$), where the statistics are much better. This trend is at odds with the 
observations of \citet{conselice08}, which show a decrease in the fraction of compact EROs 
with increasing redshift, while the proportion of peculiar and interacting galaxies increases. 
As was mentioned before, it is not clear what range of $B/T$ corresponds to irregular galaxies. 
Indeed, \citet{conselice08} observe that disks and distorted spheroids EROs have light 
concentrations that are consistent with both early and late-type classifications. 
These authors also found that although peculiar and early-type galaxies occupy quite  
different locii in the light concentration parameter space, disks and distorted spheroids 
occupy intermediate values. Thus, it is expected to be possible to separate peculiar from 
early-type galaxies using the $B/T$ parameter alone only in a sample free of disks and 
distorted spheroids. Otherwise, one would expect to find a distribution of galaxies with 
$B/T$ values varying smoothly across the full dynamical range, from 0 to 1.  The lack of 
bimodality is borne out by the smooth distribution of the predicted values of $B/T$ plotted 
in Fig.~\ref{fig:bt}.

Both Table~\ref{tab:morfo} and Fig.~\ref{fig:bt} suggest that early-type EROs are redder, 
although the dependence of the morphological mix on both magnitude range and redshift 
is stronger. As can be seen from Fig.~\ref{fig:zmedian}, the \citeauthor{bower06} model 
predicts that at a given redshift it is more likely for redder EROs to be brighter. This 
is consistent with previous predictions since brighter EROs are more likely to exhibit an 
early-type morphology. Observations show just the opposite: peculiar galaxies appear to be 
redder \citep{moriondo00,smail02,smith02}. Nevertheless, with the observations 
of \citet{moustakas04} it is clear that this result is very sensitive to the magnitude 
range adopted and the incorporation of peculiar galaxies into the late-type class. 

\subsection{Quiescent vs. Starburst EROs}\label{sec:qui}

\begin{figure}
{\epsfxsize=8.5truecm
\epsfbox[18 23 510 488]{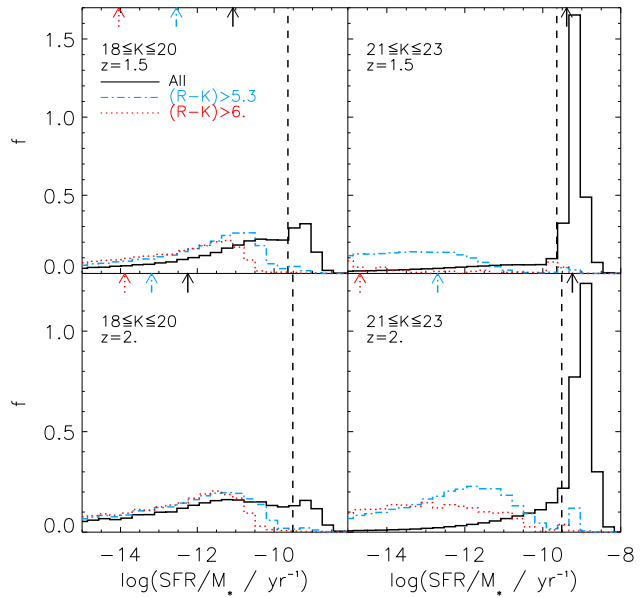}}
\caption
{The distribution of the specific star formation rate predicted by the \citeauthor{bower06} 
model, for different redshifts and 
K-band magnitude limits, as indicated in the legend within each panel. 
The arrows mark the median of the histogram drawn with the same line style. Median values for EROs in the 
upper right panel are out of the shown range.  
The vertical dashed lines indicate the inverse of the age of the
universe at the labelled redshift.
}
\label{fig:sm}
\end{figure}

We now consider the level of star formation activity in the model EROs. Galaxies with relatively 
high star formation rates can still possess the red colours associated with EROs if they are 
heavily extincted. Alternatively, the red colour could arise from an old stellar population 
with little recent star formation. 

We start by examining the specific star formation rate, $SFR/M_{*}$, that is, the star 
formation rate ($SFR$) per unit stellar mass. This ratio has dimensions of inverse time and 
is a measure of how rapidly the galaxy is adding to its stellar mass at the time of observation. 
If the specific star formation rate is comparable to the inverse of age of the universe at 
the epoch the galaxy is seen, then a substantial addition to the stellar mass is taking place. 
Fig.~\ref{fig:sm} shows the distribution of this parameter for samples of galaxies selected by 
K-magnitude and also by $(R-K)$ colour. The specific star formation rates of most galaxies are 
smaller than the inverse of the respective age of the universe at each redshift plotted, 
particularly so for the EROs. Only a small fraction of galaxies in the samples selected by K-magnitude 
alone are vigorously adding to their stellar mass. The distribution of the specific star formation 
is smooth, with no evidence for bimodality and little dependence on redshift or K-magnitude. 
The long tail of EROs with very low specific star formation rates suggests that for a large proportion 
of these objects the level of star formation activity is relatively unimportant and they 
are passively evolving, a point to which we return later on in this section. 

\begin{figure}
{\epsfxsize=8.5truecm
\epsfbox[28 8 550 478]{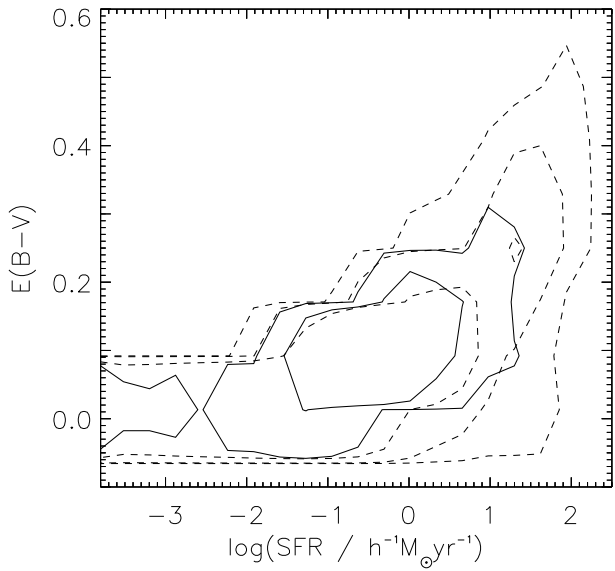}}
\caption
{The dust extinction, as quantified through $E(B-V)$ versus star formation rate ($SFR$) predicted 
by \citeauthor{bower06} at $z=1$ (solid contours) and $z=1.5$ (dashed contours) for EROs with  
$K\leq 20.$ and $(R-K)>5.3$. 
}
\label{fig:sfr_ext}
\end{figure}

Next we consider whether or not the star formation rate correlates with extinction for EROs. 
For a galaxy with a relatively high star formation rate to meet the colour criteria to be 
called an ERO, we would expect a correspondingly high level of extinction. In 
Fig.~\ref{fig:sfr_ext} we plot the extinction (see \S~2 for a brief description of the 
extinction model), as quantified by $E(B-V) = 
(B-V)_{\rm with \, dust} - (B-V)_{\rm no \, dust}$, with the filters defined in the rest frame, 
against star formation rate. Each contour represents a change in the number density of galaxies 
by a factor of 10. There is a weak trend of extinction increasing with 
star formation rate, but only for rates above $\log(SFR/h^{-1}M_{\odot}{\rm yr}^{-1})>-0.5$.
The model predicts that EROs have a wide range of extinctions, with colour 
excesses varying from $0$ to $\sim0.6$. 
A little over $\sim 10$\% of EROs have a colour excess $E(B-V)>0.1$; these are also the most 
actively star forming galaxies.  

The comparison between the values of $E(B-V)$ predicted by the model and empirical estimates 
in the literature is not straightforward. Typically, empirical estimates are made by assuming 
an extinction law and that the dust is in a screen in front of the stars and that no scattered 
light reaches the observer. As described in Section 2, in the model we assume that the dust 
and stars are mixed together and that some of the dust is in the form of molecular clouds 
\citep[see][for a discussion]{granato00}. In this case, the attenuation of starlight 
depends upon the randomly assigned inclination angle of the disk. We adopt a Milky Way extinction 
law; however, the attenuation {\it observed} in practice for a model galaxy is a function of 
the viewing angle and wavelength \citep[see fig. 11 of ][]{granato00}. For these reasons, it 
is of limited use to compare actual values of $E(B-V)$ unless the same assumptions have been made. 
We note in passing that \citet{nagamine05} were forced to fix an extreme colour excess of 
$E(B-V)=0.4$ in a foreground screen extinction model with a \citeauthor{calzetti00} extinction 
law in their gas dynamical simulations in order produce enough EROs. Here we have shown that 
with an {\it ab initio} calculation of the dust extinction, such an extreme assumption is 
not supported. 

\begin{figure}
{\epsfxsize=8.5truecm
\epsfbox[65 7 555 474]{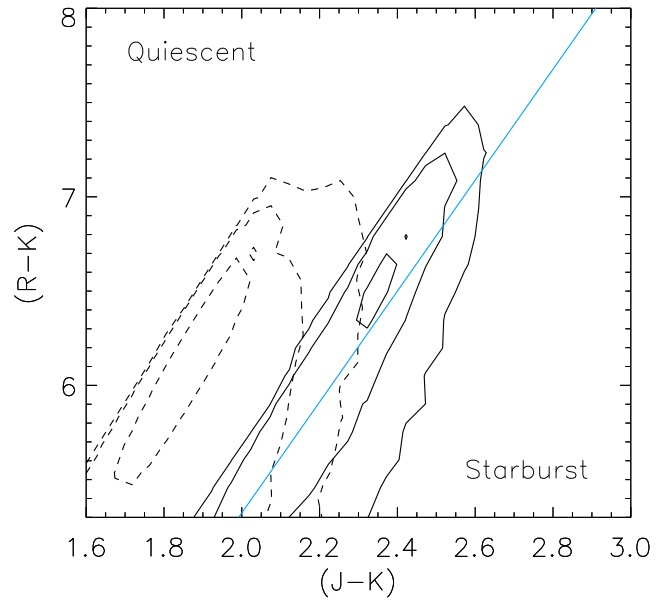}}
\caption
{The \citet{pozzetti00} colour-colour diagram for galaxies with $18\leq K\leq 20$, at 
redshift $z=2$ (solid contours) and $z=1.5$ (dashed lines) as predicted 
by the \citeauthor{bower06} model. The sloping solid line was proposed by Pozzetti \& Mannucci to divide 
the colour-colour plane into regions occupied by quiescent galaxies (to the left of the line) and 
starbursts (to the right). Each contour in the plot represents a change by a factor of 10 in the number density of galaxies.
}
\label{fig:pozzetti_class}
\end{figure}

Despite the tendency for EROs that are actively star forming to also be dustier, we find little evidence 
for any distinct bimodality. There appears to be a continuous variation in extinction and star formation 
rate, with no obvious place to separate EROs into distinct populations that are either passive or 
active and dusty. Another way to look at this issue is in the $(R-K)$ vs $(J-K)$ colour-colour plane, 
as proposed by \citet{pozzetti00}. These authors suggested that EROs with $(R-K)>5.3$ in the redshift 
range $1\leq z\leq 2$ could be classified as either {\it quiescent} or {\it starburst} galaxies 
according to their location on the $(R-K)$ vs. $(J-K)$ colour diagram (see Fig.~\ref{fig:pozzetti_class}). 
This scheme has been shown to work observationally in $\sim 80\%$ of cases 
by comparing spectral templates to galaxies on either side of the dividing line in the colour-colour 
plane \citep[e.g.][]{miyazaki03}. In the model, additional information is available to us about the 
star formation and merger history of each galaxy, such as the lookback time to the last burst of star 
formation triggered by a galaxy merger or by the collapse of an unstable disk. Hence, in addition to 
using the location of a galaxy in the colour-colour plane, we can also employ a classification based 
on the time of the most recent burst of star formation:  
\begin{itemize}
\item {\it Quiescent}: Passively evolving galaxies, whose last burst of star formation occurred more 
than $1\,$Gyr ago.\\
\item {\it Starburst}: This classification includes galaxies that are either experiencing a burst (which we will refer 
to as {\it bursty}) when they are observed or which have experienced a burst of star 
formation within the past $1$Gyr (which we shall call {\it post-burst}). 
\end{itemize}
{\it Post-burst} EROs have been distinguished observationally from passively evolving galaxies by 
\citet{doherty05}, who found that almost $40\%$ of the EROs dominated by an old stellar 
population had spectral indications of the occurrence, within the last $\sim1\,$Gyr, of 
a secondary episode of star formation, in which a small percentage, $\sim 7 \%$, of the total 
galactic mass was formed. 

\begin{figure}
{\epsfxsize=8.5truecm
\epsfbox[65 7 555 474]{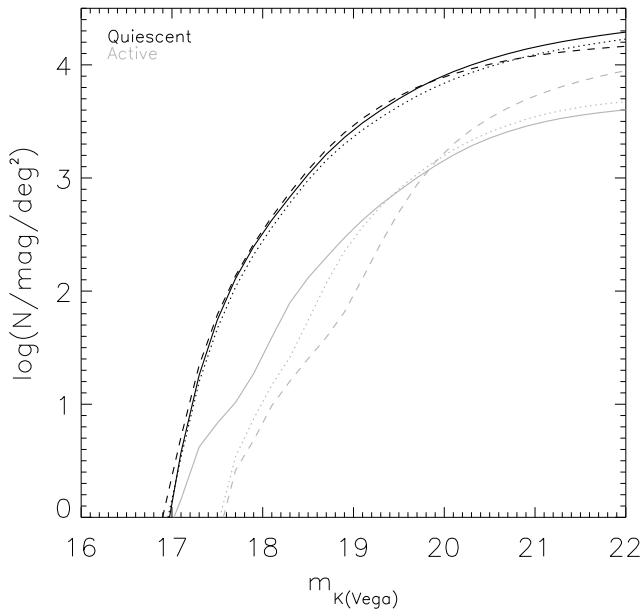}}
\caption
{The number counts of  EROs with $(R-K)>5.3$, as predicted by the \citeauthor{bower06} model, split into 
two populations. The black lines in each case show the counts for the ``quiescent'' part 
of the population.  The different grey lines refer to ``active'' population defined 
according to: 1) the time since the last episode of star formation triggered by a 
galaxy merger or the collapse of an unstable disk (solid lines); this is our standard definition 
of a starburst galaxy; 2) whether the galaxy occupies the ``starburst'' locus in the Pozzetti \& Mannucci 
color-colour diagram (dashed lines - see Fig.~\ref{fig:pozzetti_class} diagram), or 3) the magnitude of 
the specific instantaneous star formation rate (dotted lines - see text for further details). 
}
\label{fig:pozzetti}
\end{figure}

We compare in Fig.~\ref{fig:pozzetti} the cumulative number counts of EROs predicted by
the \citeauthor{bower06} model, classified as {\it quiescent} or {\it starburst} using the two methods described 
above i.e. the location in the colour-colour plane, or the occurrence of the last burst of star 
formation. Fig.~\ref{fig:pozzetti} shows that the counts of {\it quiescent} EROs as defined 
by their colour or the time since the last starburst, agree spectacularly well. This population 
dominates over the active or dusty starburst EROs. In the active case, the counts corresponding 
to the different definitions are not in such good agreement, so it is hard to conclude that the alternative  
definitions of active galaxies pick out similar objects. In one sense our definition of starburst 
is flawed in that quiescent galaxies could be experiencing similar star formation rates in their 
galactic disks, with the only distinction being that this star formation was not triggered by a 
galaxy merger or an unstable disk. 

We have also investigated a third scheme to divide model EROs into active and passive populations, 
based on the magnitude of the specific star formation rate, i.e. the instantaneous star formation 
rate divided by the stellar mass of the galaxy. The specific star formation rate quantifies the 
contribution of the current episode to the total stellar mass. A high specific star formation rate 
means that the current star formation will have a observable impact on the overall galaxy colour 
or on the visibility of line emission against the stellar continuum. With this definition, the 
classification into active and passive does not depend upon some triggering event, such as a galaxy merger 
or the dissolution of an unstable disk. This means that the active ERO population could be made up 
of ``quiescent'' disks or vigorous starbursts. In Fig.~\ref{fig:pozzetti}, we find similar results to 
those obtained in the two classification schemes discussed above if we place the division at a specific 
star formation rate of $SFR/M_{*}=10^{-11}{\rm yr}^{-1}$. 

\begin{table}
\caption
{
The predicted mix of {\it starburst:quiescent} galaxies (see text for definitions) in the \citeauthor{bower06} model, 
for K-selected galaxy samples and for EROs with $18\leq K \leq 20$ at $z=1,\, 1.5, \, 2$. 
EROs are defined with different colour cuts, as indicated.   
}
\begin{center}
\begin{tabular}{ccccc}
\hline
Classification  &  \multicolumn{4}{c}{$z$}  \\
\cline{2-5}
Scheme & $(R-K)$ & 1.0 & 1.5 & 2.0 \\
\hline
                    & All & 18:82 & 24:76 & 39:60 \\
Burst               & 5.0 & 1:99  & 12:88 & 24:76 \\
occurrence          & 5.3 & 0:100 & 10:90 & 21:79 \\
                    & 6.0 &  -    &  5:95 & 12:88 \\
\hline
Colour-Colour       & All & 34:64 & 20:80 & 43:57 \\
Pozzetti \&         & 5.3 & 15:85 &  1:99 & 25:75 \\
Mannucci (2000)     & 6.0 &   -   &  0:100& 14:86 \\
 \hline
\end{tabular}\label{tab:s_morfo}
\end{center}
\end{table}

Table~\ref{tab:s_morfo} summarizes the predicted percentages of {\it quiescent} and {\it starburst} 
galaxies for K-selected samples and for EROs, according to the classification methods outlined above. 
The magnitude range has been chosen to be close to that used in observations. Spectroscopic 
studies of EROs, with $(R-K)>5$ or $5.3$ and magnitude limits around $K\sim20$ \citep[e.g][]
{smail02,simpson06}, find roughly equal numbers of passively evolving and dusty star forming EROs. Similar ratios where found in observational studies \citep[e.g.][]{mannucci02,cimatti03} that used the Pozzetti \& Mannucci color-colour diagram to distinguish between  {\it quiescent} and {\it starburst} EROs.
Therefore, the predicted mix of {\it quiescent} and {\it starburst} EROs does not match the 
observations.  The \citeauthor{bower06} model underpredicts the number of dusty star forming EROs. 
This is likely to be due to the predicted scarcity, $\leq 1\%$, of {\it bursty} galaxies among EROs. 
This result can be connected with the number counts and redshift distribution of EROs, which 
suggest the need to review the star formation at $z>1$ in the \citeauthor{bower06} model. 
Either the star formation is quenched too soon due to a lack of cold gas or the burst timescale 
is too short. 

Table~\ref{tab:s_morfo} shows the tendency for the percentage of galaxies classified by the 
lookback time of the burst as {\it quiescent} to fall with increasing redshift. This matches 
the expectation for young galaxies to become the dominant population at high redshift. 
Nevertheless, this tendency is not as clear when the classification is made following 
the \citet{pozzetti00} colour-colour method. 

No clear dependency is found in the ratio between {\it quiescent} and {\it starburst} 
for K-selected galaxies with magnitude range. 
The same can be said for EROs classified with the \citet{pozzetti00} method. However, 
the percentage of {\it quiescent} EROs, defined by the occurrence of their last burst, 
increases by $\sim 12$\% from a bright magnitude range, $18\leq K \leq 20$, to a faint one, 
$21\leq K \leq 23$. \citet{cimatti02a} observed the opposite tendency, though with a smaller 
variation (a $\sim 4\%$ change). Comparing observations from studies with different magnitude 
limits \citep[][]{roche02,cimatti02a,cimatti03,miyazaki03,vaisanen04} no clear tendency is seen. 
\citet{smith02} proposed that the change in the slope of ERO number counts evident 
in Fig.~\ref{fig:erodndm} is due to a change in the nature of bright and faint EROs. 

From Table~\ref{tab:s_morfo} there is a clear tendency for the reddest EROs to be dominated 
by quiescent galaxies, independent of the classification method. The same tendency was found 
by \citet{simpson06}, whose sample of EROs, was classified into old and starburst using the 
\citet{pozzetti00} method.

Finally, following the scheme set out by \citet{malbon07}, the models trace the 
growth of supermassive black holes, so in principle it is possible to identify if an ERO also has 
some sort of nuclear activity. The mass of the black hole may grow through mergers of pre-existing black holes, the 
accretion of cold gas during a starburst or the accretion of cooling flow gas in a quasi-static 
halo. The accretion of cold gas during mergers is usually associated with a QSO phase and the 
feeding of the black hole from the cooling flow with a ``radio mode'' of activity \citep{croton06}. 
The timescale for the ``radio mode'' activity is ill-defined in the model. Some fraction of the mass 
that would be involved in a cooling flow is diverted onto the black hole and the energy released 
stifles the cooling. This is assumed to operate over the lifetime of the halo.

In the case of QSO activity, the timescale for the accretion of cold gas onto the black hole is specified more 
transparently, but is poorly constrained. The choice of timescale has an impact on the luminosity 
of the QSO and so is chosen to reproduce the QSO luminosity function \citep{malbon07}. 
With the current choice of parameters for these timescales, none of the EROs modelled by \citeauthor{bower06} show 
QSO activity. 
Observations suggest that the presence of AGNs among EROs is rare, and when present they are 
likely to be in a weak phase, rather than in a QSO state. \citet{roche02} detected $5$ ($16\%$) 
EROs with radio emission that was compatible with a weak AGN, and $1$ ($3\%$) were detected in X-ray,  
which suggested an AGN in a stronger phase. \citet{smail02} estimated that $6\%$ of their radio 
detected EROs were AGNs.

\section{Other properties of EROs}
\label{sec:mass}

\begin{figure}
{\epsfxsize=8.5truecm
\epsfbox[53  10 568 474]{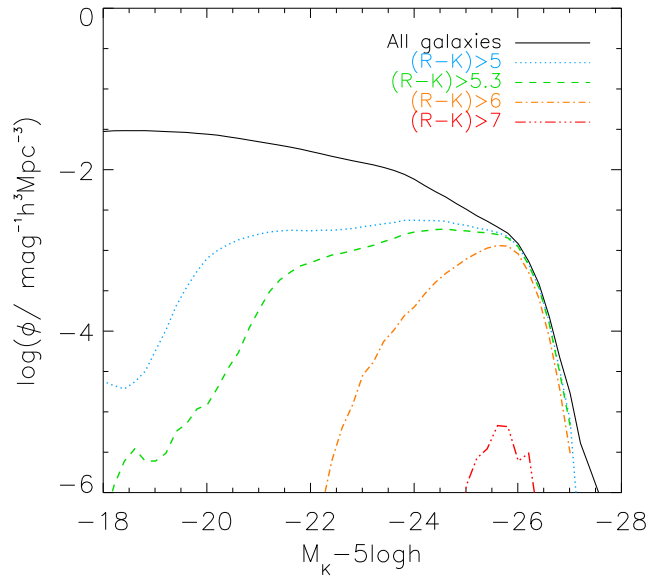}}
\caption
{
The K-band luminosity function in the observer's frame predicted by the \citeauthor{bower06} model at redshift $z=1.5$, for all galaxies (solid line) and EROs redder than $(R-K)=5$ (dotted line), $5.3$ (dashed line), $6$ (dash-dotted line), and $7$ (dash-triple dotted line). 
}
\label{fig:bower_lf}
\end{figure}

In this section we explore the model predictions for basic properties of EROs, such 
as luminosity, stellar mass, host halo mass, galaxy size and age.  
Given that the \citeauthor{bower06} model gives the best match to the observed counts 
of EROs, we shall again only show predictions from this model in this section. 

\subsection{Luminosity} 
Fig.~\ref{fig:bower_lf} shows the predicted $K-$band luminosity function 
at $z=1.5$, for all galaxies and for different samples of EROs defined 
by $(R-K)$ colour. \citeauthor{bower06} showed that the luminosity function of all galaxies 
in the $K$-band agrees well with the available observations up to $z \sim 1.5$. 
Fig.~\ref{fig:bower_lf} shows that faintwards of $L_{*}$, only a fraction of 
galaxies have the colour required to be classified as an ERO. Furthermore 
this fraction falls dramatically with declining luminosity. The fraction of 
faint galaxies that are EROs also drops significantly as the $(R-K)$ colour 
threshold gets redder. Brightwards of $L_*$, however, essentially all galaxies 
are predicted to be EROs, until $(R-K)>7$ is reached. This result matches the 
observations of \citet{conselice08}, who found the reddest galaxies at $z\geq1$ 
to be also the brightest in K-band.

\begin{figure}
{\epsfxsize=8.5truecm
\epsfbox[53  10 568 474]{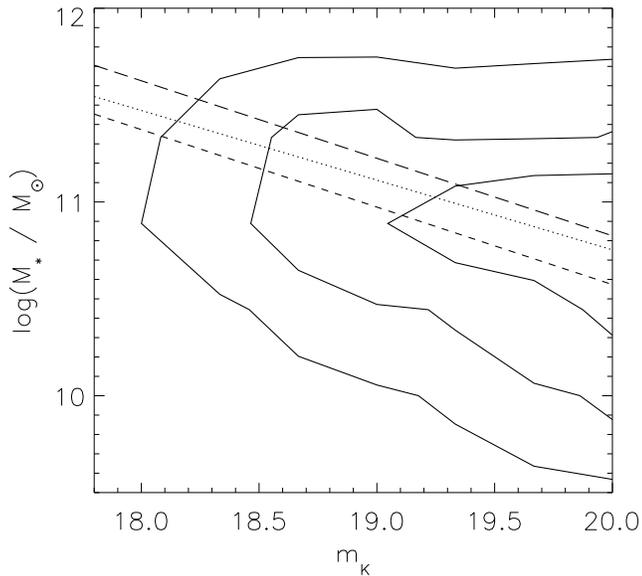}}
\caption
{
The number density of \citeauthor{bower06} model galaxies with $K < 20$ at $z=2$ in the stellar mass - $K-$band 
apparent magnitude plane (no colour selection is applied here). The dotted line shows the 
best fit stellar mass for the model. The two dashed lines show different estimates of 
the stellar mass of K20 survey galaxies made by \citet{fontana04}, after applying 
an approximate correction to account for differences in the stellar initial mass function 
adopted in the two calculations. 
}
\label{fig:mass_k}
\end{figure}

\subsection{Stellar mass} 

We first compare the predicted stellar masses in a $K$-selected sample 
with masses estimated from the K20 survey. The contours in Fig.~\ref{fig:mass_k} 
show the distribution of stellar mass for \citeauthor{bower06} galaxies with $K <20$ at z=2. Each contour represents a change in the number density of galaxies 
by a factor of 10.
The dotted line shows a best fit to the stellar mass - $K$-band apparent magnitude 
relation. The dashed lines show the best fit to the mass estimated for galaxies 
in the same magnitude range around $z \sim 2$ in the K20 survey \citep{daddi04}. 
The two estimates correspond to the best fits when using the full photometry (lower 
line) or just one colour ($(R-K)$, termed the maximal mass estimate). Note that these curves 
have been shifted to account for the different stellar initial mass function (IMF) 
adopted by \citet{fontana04}. \citeauthor{bower06} adopt a \citet{kennicutt_imf} IMF, whereas 
\citeauthor{fontana04} use a \citet{salpeter55} IMF. The difference in mass to light 
ratio depends on star formation history and metallicity. \citeauthor{fontana04} 
give some examples of how the mass to light ratios differ for these two choices of IMF. 
For a stellar population with age $<1$Gyr, they state that the mass to light ratio 
is a factor $\sim 1.4$ higher when using a \citeauthor{salpeter55} IMF compared 
with \citeauthor{kennicutt_imf}; the difference becomes a factor 2.2 for older populations. 
We have applied an indicative correction, dividing the masses inferred with a \citeauthor{salpeter55} IMF by a factor 1.7. The stellar masses estimated by 
\citeauthor{fontana04} for K20 galaxies are in very good agreement with the model predictions. 

\begin{figure}
{\epsfxsize=8.5truecm
\epsfbox[35 23 521 488]{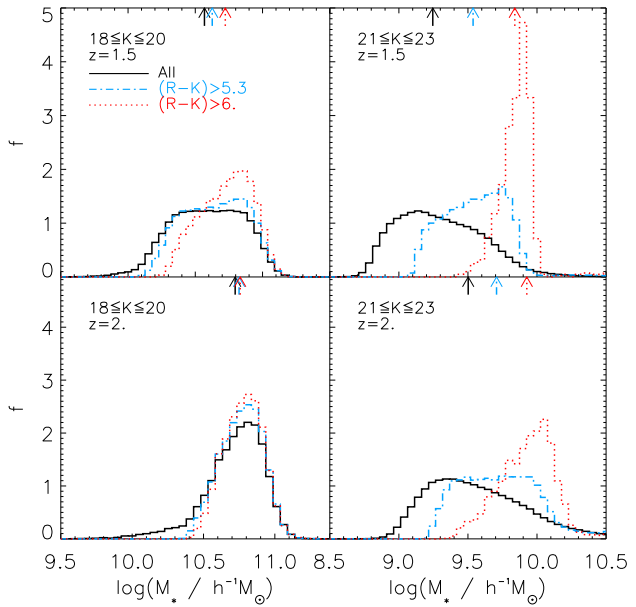}}
\caption
{The distribution of stellar mass in the \citeauthor{bower06} model for galaxies at $z=1.5$ (upper panels) 
and $z=2$ (lower panels) in the magnitude ranges $18\leq K \leq 20$ (left panels) 
and $21\leq K \leq 23$ (right panels). In each panel, the distribution of galaxies 
at the corresponding redshift and magnitude range without any further selection is 
shown as a solid line histogram. The distribution of EROs with $(R-K)>5.3$ is plotted as a dash-dotted line histogram; the dotted histogram shows the distribution of EROs with $(R-K)>6$. Median stellar mass values for each distribution are indicated 
by an arrow with the same line style as the corresponding histogram.
}
\label{fig:mass}
\end{figure}

We now compare the stellar masses of samples defined by a $K$-band selection and a 
$(R-K)$ colour threshold. 
Fig.~\ref{fig:mass} shows the stellar mass distribution of galaxies at 
redshifts $z=1.5$ and $2$, in two magnitude bins: $18\leq K\leq 20$, 
which probes the region around $L_*$ in the luminosity function and 
$21\leq K\leq 23$, which is sensitive to the faint end of the luminosity 
function at these redshifts. Fig.~\ref{fig:mass} clearly shows 
the tendency for redder galaxies to be more massive, independent of the redshift 
or magnitude range. This distinction in the mass of EROs is more pronounced for 
the fainter magnitude range, due to the smaller fraction of EROs. 

At $z=1.5$, the \citeauthor{bower06} model predicts that galaxies with 
$21\leq K\leq 23$ have a median stellar mass of $\sim1.8\times10^{9}\, h^{-1}M_{\odot}$, 
while the median mass of the subset of these galaxies which are also redder 
than $(R-K)=5.3\,(6)$ is appreciably higher, $\sim3.6\times10^{9}\, 
h^{-1}M_{\odot}\,(\,\sim6.8\times10^{9}\, h^{-1}M_{\odot})$. Therefore, EROs 
are predicted to be the most massive galaxies present at the time. \citeauthor{bower06} predicts that EROs are indeed the most 
massive galaxies in the redshift range $1\leq z\leq 2$. In particular, at $z=1.5$, 
EROs with $(R-K)>5.3$ and $K\leq19.7$ account for $\sim 74\%$ of all  
galaxies with stellar masses in excess of $M_{*}>10^{11}h^{-1}M_{\odot}$, 
in remarkably good agreement with the fraction found observationally 
by \citet{conselice08}. This percentage increases with redshift. 

The stellar mass distributions at higher redshift extend to larger masses 
than those at lower redshift. This arises because we are using a fixed 
apparent magnitude bin and simply sample intrinsically more luminous, 
and consequently more massive galaxies at high redshift.

\subsection{Galaxy radii} 

\begin{figure}
{\epsfxsize=8.5truecm
\epsfbox[31  10 559 474]{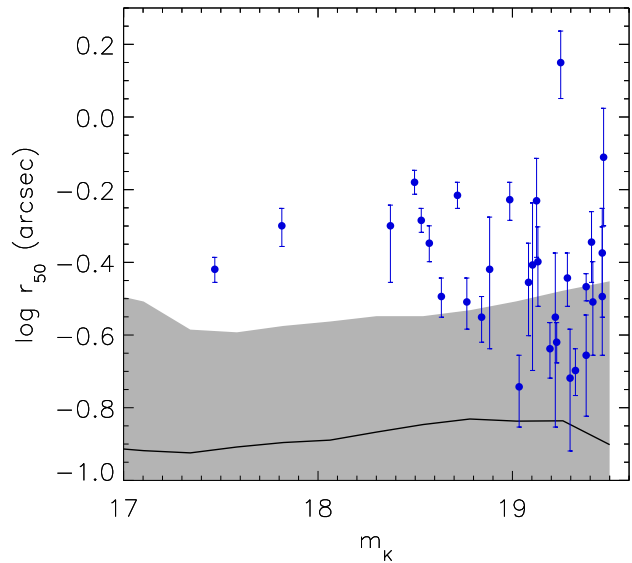}}
\caption
{The median galaxy radius as a function of apparent magnitude predicted by the \citeauthor{bower06} 
model for EROs with $(R-K)>5$ and $K<19.5$. The shaded region shows the $90$-percentile range for predicted radii. For comparison, observational data for EROs 
with the same selection from \citet{roche02} are shown as filled circles.  
}
\label{fig:r50}
\end{figure}

{\tt GALFORM} predicts the size of the disk and bulge components of galaxies by 
tracking the angular momentum of the gas which cools to make a galactic disk, and, 
in the case of mergers, by applying the conservation of energy and the virial theorem.
In Fig.~\ref{fig:r50} we plot the predicted median half-mass radius of EROs as a function of 
apparent magnitude, and compare this with observational determinations of the radii of 
EROs from \citet{roche02}. The shaded region in Fig.~\ref{fig:r50} shows the 90-percentile 
range of the model predictions. Whilst there is some overlap between the model predictions 
and the observational estimates at faint magnitudes, the model galaxies are on the whole 
too small by around a factor of two or more. It is possible that the radii of some of 
the observed galaxies may be overestimated, due to the ERO being associated with two galaxies 
which are in the process of merging, whereas the model predictions refer to the size of 
the merger remnant or to the sizes of the progenitor galaxies. In 
the \citeauthor{bower06} model we find that $\sim 8\%$ of the EROs present at $z=1.5$ 
experienced a merger in the preceeding 1 Gyr. If this fraction of EROs are considered 
as close pairs in the process of merging and are assigned larger sizes (by a factor 2), 
then the predicted $90$-percentile range would cover most of \citet{roche02} data points.

\citet{almeida07} also reported that bright, local early-type galaxies are observed 
to have larger radii than predicted by the model. The solution to this problem is unclear. Semi-analytical 
models assume that the angular momentum of the infalling gas is conserved whereas 
numerical simulations of disk formation show that this is not always the 
case \citep[e.g.][]{okamoto05}. Loss of angular momentum would make the problem even 
worse. The self-gravity of the baryons and their pull on the dark matter halo make the 
disk and bulge components smaller; \citet{almeida07} showed that if this contraction 
of the dark matter halo in response to the presence of the galaxy could somehow be 
switched off, the correct slope is predicted for the radius-luminosity relation.   
 
\subsection{Dark halo mass}
\label{sec:mh}
\begin{figure}
{\epsfxsize=8.5truecm
\epsfbox[18 23 508 488]{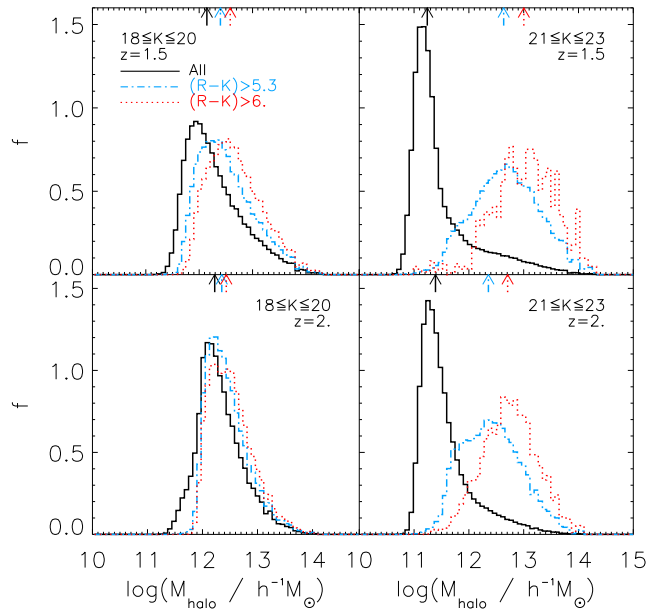}}
\caption
{
The distribution of host dark halo mass predicted in the 
\citeauthor{bower06} model at redshift $z=1.5$ (upper panels) 
and $z=2$ (lower panels) for galaxies in the magnitude ranges 
$18\leq K \leq 20$ (left panels) and $20\leq K \leq 23$ (right panels). 
In each panel, the distribution for galaxies at the corresponding redshift 
and magnitude range but without any further selection is shown as a solid line histogram. 
The distribution of halo masses for EROs with $(R-K)>5.3$ and with $(R-K)>6$ are plotted with a dash-dotted and a dotted line histogram, respectively. Median halo mass values for each distribution are 
indicated by an arrow with the same line style as the corresponding histogram.
}
\label{fig:mh}
\end{figure}

We have already shown that EROs tend to be the most massive galaxies in place 
at a given redshift. One might naturally expect therefore that they should be 
hosted by the most massive dark matter haloes present at a given epoch. In practice, 
the efficiency of galaxy formation tends to drop with increasing halo mass, as 
revealed observationally by an increase in the mass to light ratio in clusters compared with 
galactic haloes \citet{eke04}, and so the trend between the 
luminosity of the main galaxy within a halo and halo mass could be quite weak. 
Furthermore, we have also seen that EROs can have a wide range of luminosities and 
presumably can be hosted by a wide range of halo masses. Fig.~\ref{fig:mh} shows  
the distribution of dark halo mass for haloes which host galaxies in two ranges 
of $K-$band magnitude at $z=1.5$ and $z=2$. It is clear how these 
distributions shift to larger masses when a cut on $(R-K)$ colour is also applied. 
This effect is very pronounced at $z=2$, where the median mass of haloes hosting 
EROs is ten times larger than that of a sample of galaxies without a colour selection. 
This shift in host halo mass and its dependence on the redness of the (R-K) cut will 
have consequences for the predicted clustering of EROs. This is addressed further 
in Paper II.

\subsection{Age of stellar populations}
\label{sec:age}

\begin{figure}
{\epsfxsize=8.5truecm
\epsfbox[18 25 508 488]{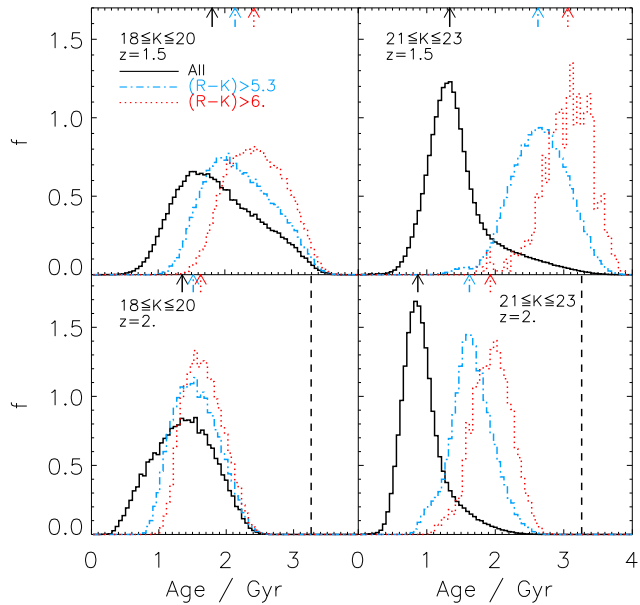}}
\caption
{
The distribution of the rest-frame V-band 
luminosity weighted stellar age predicted by the Bower et~al. model at redshift $z=1.5$ (upper panels) 
and $z=2$ (lower panels) for galaxies in the magnitude ranges 
$18\leq K \leq 20$ (left panels) and $20\leq K \leq 23$ (right panels). 
In each panel, the distribution for galaxies at the corresponding redshift 
and magnitude range but without any further selection is shown as a solid line histogram. 
The distribution of ages for EROs with $(R-K)>5.3$ and with $(R-K)>6$ is plotted with a dash-dotted and a dotted line histogram, respectively. Median ages for each distribution are 
indicated by an arrow with the same line style as the 
corresponding histogram. The vertical dashed lines show the age of the 
Universe at z=2 (at $z=1.5$ the corresponding lines lie to the right 
of the graph range).
}
\label{fig:age}
\end{figure}

The formation histories of galaxies found in the deepest gravitational potential 
wells are shifted to earlier times compared with galaxies of the same mass found 
in less extreme haloes. This is a natural consequence of hierarchical structure 
formation \citep{neistein06}. The most massive haloes will 
tend to be found in regions with higher than average overdensity. Drawing an 
analogy with the spherical collapse model, the evolution of such a patch of 
universe will be accelerated with respect to an average density patch, with 
the consequence that haloes, and hence galaxies, will start to form 
earlier. This will lead to galaxies in massive haloes having older stellar 
populations. A further difference in the age of galaxies in different mass 
haloes will be introduced if there is a physical process which acts to 
suppress galaxy formation preferentially in the more massive haloes. 

The predicted rest-frame V-band luminosity weighted age of the stellar populations 
is plotted in Fig.~\ref{fig:age}. The model EROs ages have a wide distribution, 
with medians in the range $\sim [1.5,3]$ Gyr. 

Few observational studies have estimated the age of EROs. Using pure 
luminosity evolution models, \citet{vaisanen04} estimated a formation 
redshift of around $z_{f}\sim3$ for EROs. The average observed redshift 
of their sample is approximately $z=1$, thus implying an age of $\sim 3\, $Gyr. 
At $z=1$, the \citeauthor{bower06} model predicts that EROs with $(R-K)>5.3$ have a median age of 
$\sim 3\, $Gyr, for both the bright and faint magnitude ranges in Fig.~\ref{fig:age}, 
in good agreement with the estimate of \citet{vaisanen04}. 
\citet{cimatti02a} compared the average spectrum of quiescent EROs with stellar 
population synthesis models, and derived a lower limit to the age of EROs of 
$\sim 3\, $Gyr. Within a similar magnitude and redshift range to those of the  
\citet{cimatti02a} sample, the \citeauthor{bower06} model predicts that quiescent EROs have 
an average age of $3.45\, $Gyr, in agreement with the observational estimate. 
Here, as in section \S\ref{sec:qui}, we consider quiescent galaxies as those 
that have not experienced a recent burst of star formation.

Proceeding in a similar way to \citet{cimatti02a} but considering a sample of 
mainly dusty starburst EROs, \citet{schaerer07} found ages of $\sim 0.6\, $Gyr. 
The mean redshift of these EROs is $z=1.5$. At this redshift, simulated EROs with 
$(R-K)>5.3$ and $18\leq K\leq 20$, experiencing a starburst are expected 
to have a mean age of $2.4\, $Gyr. The observed sample is composed of only 8 EROs, 
found behind lensing clusters. Thus, it is quite possible that the observational 
results are not representative of the averaged age of bursting EROs, but rather 
are extreme cases. More observations will be needed to reach firm conclusions 
about the age of bursty EROs.

\section{Conclusions} \label{sec:conclusions}

In this paper we have extended the tests of the {\tt GALFORM} galaxy formation 
code to include red galaxies at $z>1$. \citet{almeida07lrg} started this series 
of comparisons by presenting model predictions for the abundance of luminous 
red galaxies at lower redshifts, $z=0.24$ and $z=0.5$. The EROs we 
consider in this paper are not, as a whole, necessarily as intrinsically bright 
as luminous red galaxies. Nevertheless, the much longer look-back time to $z>1$ 
poses challenges if hierarchical models are to accommodate EROs. EROs are thought 
to be made up of relatively old passively evolving stellar populations and dusty 
starbursts \citep[e.g.][]{mccarthy04}. The former implies a high formation redshift for 
the stars and the latter suggests the presence of dust-enshrouded objects with 
high star formation rates.  

An important feature of the calculations presented here is that they are readily 
connected to observations. The semi-analytical model applies simple physical 
recipes to determine the fate of the baryonic component of the universe. The output 
is the full record of galaxy mergers and the star formation history for a wide 
range of galaxies. 
The composite stellar spectrum predicted for each galaxy can then be convolved 
with filter transmission curves so that samples of galaxies can be constructed 
with the same photometric selection as the data. Moreover, by using a chemical 
evolution model to track the metallicity of the cold gas in galactic disks and by 
computing the size of the disk and bulge components, the model is able to calculate 
the amount of extinction experienced by starlight at any wavelength. This is a particularly 
important consideration given our findings regarding the nature of EROs. These galaxies seem to 
have a range of properties so it would be incorrect to take short cuts and assume 
that EROs are exclusively starburst galaxies or galaxies with little recent star formation, 
and then to try to use this as a proxy instead of applying the proper photometric 
selection of EROs. 

We have tested two published models of galaxy formation, those of \citet{baugh05} and  
\citet{bower06}. The parameters in both models were set to reproduce observations 
of the local galaxy population, though with different emphasis on which observations 
were the most important to reproduce closely. Both models enjoy successes in matching 
observations of the high redshift universe. The \citeauthor{baugh05} model reproduces the number 
counts of sub-mm selected galaxies and the luminosity function of Lyman-break galaxies, 
whereas \citeauthor{bower06} model matches the evolution of the K-band luminosity function 
and the inferred evolution of the stellar mass function. 

The \citeauthor{baugh05} model, despite the aforementioned successes at high redshift, 
underpredicts the counts of EROs by an order of magnitude, mirroring the predictions 
of earlier semi-analytical models \citep[see e.g.][]{smith02}. The model predicts some 
passively evolving EROs (around $1/5^{\rm th}$ of the number it should do, depending 
on the magnitude) but does  not produce any dusty starbursts with the colour of EROs. 
There are dusty starbursts in  the \citeauthor{baugh05} model, as these are the sub-mm 
sources which match the observed SCUBA counts. However, there seems to be little overlap 
between the population of sub-mm galaxies and EROs in this model \citep[see][for the observational view on the connection between these two types of galaxy]{smail99}. 

The \citeauthor{bower06} model, on the other hand, gives an impressively close match to the 
number counts of EROs. If anything, this model predicts somewhat too many galaxies with 
the colours of passively evolving stellar populations at high redshift; intrinsically 
red galaxies dominate over dusty starbursts at all magnitudes in this model.  

What does this tell us about the physics of massive galaxy formation? We 
experimented with the \citeauthor{baugh05} model to see if its predictions could be reconciled 
with the observed number of EROs on changing the model parameters. An obvious 
place to start was the duration of starbursts. In the \citeauthor{baugh05} model, starbursts 
have a long duration to prevent the dust from getting too hot, which would reduce the 
counts of sub-mm galaxies. Reducing the duration of starbursts made little difference 
to the predicted ERO counts, again suggesting that dusty starbursts are not the dominant 
population of EROs. The key seems to be that \citeauthor{bower06} model gives a better match 
to the observed evolution of the K-band luminosity function, which means that this model 
puts massive galaxies in place earlier than in the \citeauthor{baugh05} model. This problem has 
been revealed from a different point of view by \citet{swinbank08}, who argued that
the stellar masses of sub-mm galaxies are too small in \citeauthor{baugh05} model. This 
difference between the two models arises from the different redshift dependence of the feedback 
processes which suppress the formation of massive galaxies and from the choice of the star formation 
timescale. In both models, a physical process operates to reduce the cooling rate in massive haloes. 
In the \citeauthor{baugh05} model this is achieved by the ejection of gas in a superwind, which 
lowers the effective baryon fraction in massive haloes, thereby reducing the rate at which gas can cool. 
In the \citeauthor{bower06} model, the feeding of a central supermassive black hole releases energy 
which stalls the cooling flow completely. The lack of a dribble of cold gas from which to form even 
a small amount of stars helps galaxies in the \citeauthor{bower06} model to attain the colours 
of EROs. In \citeauthor{bower06}, the star formation timescale scales with the local dynamical time, whereas the 
scaling is independent of redshift in the \citeauthor{baugh05} model. This means that a given amount of cold gas will 
be turned into stars more quickly at high redshift in the \citeauthor{bower06} model than in the \citeauthor{baugh05} model. 

In the \citeauthor{bower06} model, EROs are predominantly passively evolving galaxies. However, we do not 
find two distinct populations of objects, which suggests a transformation from a dusty 
starburst phase, which lasts a comparatively short time, to a longer lived quiescent 
phase. This is slightly at odds with observations which suggest a more equal split between 
passive galaxies and dusty starbursts. Also, the redshift distribution of passive galaxies 
predicted by the \citeauthor{bower06} model is more extended than is observed. This suggests that the 
star formation in massive objects may have been quenched too efficiently by the radio-mode 
AGN feedback. \citeauthor{bower06} predict that EROs are mainly spheroid dominated, though we also find 
disk dominated EROs. The \citeauthor{bower06} model predicts that EROs are the most massive galaxies in 
place at $z >1$, in agreement with observations. The main discrepancy between the 
predicted properties of EROs and observations lies with the scale sizes of EROs, which are 
smaller than observed \citep[see also][]{almeida07}.

The \citeauthor{bower06} model has proven to be successful at reproducing 
the abundance and general properties of EROs at $z\sim 1$, and also the luminosity 
function of LRGs at $z=0.24$ \citep{almeida07lrg}. Both EROs and LRGs are massive, 
bright galaxies, dominated, at least in the model, by old stellar populations. 
However, the \citeauthor{bower06} model does not reproduce the observed counts of sub-mm 
galaxies \citep{swinbank08}, which are also massive galaxies, with stellar mass 
$M>10^{11}M_{\odot}$ at $z\sim 2$, experiencing active star formation. On the other hand, 
the \citeauthor{baugh05} model does reproduce the counts and redshift distribution 
of sub-mm galaxies, but does less well at matching the abundance of red galaxies at 
$z<1$. In a later paper in this series, we will investigate the nature of the objects 
selected by different colour and magnitude criteria. The semi-analytical model is 
ideally suited to connecting galaxies identified at high redshift with their local, 
$z=0$ counterparts. We will address the issue of what fraction of today's galaxies 
had a progenitor which passed the criteria to be identified as a red galaxy and 
we will determine what fraction of the present day stellar mass was already in place 
by this epoch. 

This is the first paper in a series which examines the properties and nature of red 
galaxies in hierarchical models. Here we have presented predictions for the abundance 
and redshift distribution of EROs, along with some basic properties, such as stellar 
mass and host halo mass. In the second paper we present predictions for the clustering 
of EROs and in the third we compare different colour cuts used to select red galaxies 
and compare the properties of their present day descendants.

\section*{ACKNOWLEDGMENTS}
{
We thank S. Foucaud, C. Simpson and I. Smail for providing their measured EROs 
number counts in a table format. We also acknowledge F. J. Castander, J. Helly, A. Benson, 
R. Malbon and M. Swinbank for discussions and comments. We thank the referee for a helpful 
report. VGP acknowledges support from CSIC/IEEC, the 
Spanish Ministerio de Ciencia y Tecnolog\'{i}a and travel support from a Royal 
Society International Joint Project grant. CMB was supported by the Royal Society. 
CGL is supported in part by a grant from the Science and Technology Facilities 
Council. CA gratefully acknowledges a scholarship from the FCT, Portugal.
}

\vspace{-0.7cm}

\bibliographystyle{mn2e}


\end{document}